\documentclass[12pt]{article}

\usepackage{amsfonts, amsmath}

\usepackage[margin=1in]{geometry}
\usepackage{booktabs}
\usepackage{dcolumn}
\usepackage{setspace}
\usepackage{verbatim}
\usepackage{fullpage}
\usepackage{epsfig, graphicx,caption,mwe}
\usepackage{subcaption}
\usepackage{placeins}
\usepackage{wrapfig}
\usepackage[flushleft]{threeparttable}
\usepackage{lscape}
\usepackage{rotating}
\usepackage{epstopdf}
\usepackage{verbatim}
\usepackage{url}
\usepackage{floatrow}
\usepackage{dirtytalk}
\usepackage{appendix}
\usepackage{multirow}
\usepackage{threeparttable}
\usepackage{graphicx}
\usepackage{bm}
\usepackage{hyperref}
\hypersetup{
    colorlinks,
    linkcolor=blue,
    filecolor=blue,      
    urlcolor=blue,
    citecolor={blue},
}
\usepackage{longtable}

\usepackage{pdflscape} 

\usepackage{changepage}

\usepackage{float}

\usepackage{threeparttable}
\usepackage{longtable}

\usepackage[american]{babel} 
\usepackage{csquotes} 

\usepackage{url}

\usepackage[numbers]{natbib}
\usepackage[resetlabels,labeled]{multibib}
\newcites{A}{Appendix References}

\usepackage{array}

\usepackage{titlesec}

\titleformat{\paragraph}
{\normalfont\normalsize\bfseries}{\theparagraph}{1em}{}
\titlespacing*{\paragraph}
{0pt}{3.25ex plus 1ex minus .2ex}{1.5ex plus .2ex}

\begin{document}

\pagestyle{plain}

\newcommand{\n}{\noindent}

\def\citeapos#1{\citeauthor{#1}'s (\citeyear{#1})}

\newcommand{\tit}{{Data marketplaces can increase the willingness to share social media data at low prices}}

\title{\tit} \author{Meysam Alizadeh\thanks{\scriptsize{Department of Political Science, University of Zurich, Switzerland.  Email: \texttt{\href{mailto:alizadeh@ipz.uzh.ch}{alizadeh@ipz.uzh.ch}.} Corresponding Author.}}, Fabrizio Gilardi\thanks{\scriptsize{Department of Political Science, University of Zurich, Switzerland. Email: \texttt{\href{mailto:gilardi@ipz.uzh.ch}{gilardi@ipz.uzh.ch}.}}}} \date{\small{This draft: \today} 
}
\maketitle 

\begin{abstract}

Living in the Post-API age, researchers face unprecedented challenges in obtaining social media data, while users are concerned about how big tech companies use their data. Data donation offers a promising alternative, however, its scalability is limited by low participation and high dropout rates. Research suggests that data marketplaces could be a solution, but its realization remains challenging due to theoretical gaps in treating data as an asset. This paper examines whether data marketplaces can increase individuals' willingness to sell their X (Twitter) data package and the minimum price they would accept. It also explores how privacy protections and the type of data buyer—researchers vs. companies—may affect these decisions. Results from two pre-registered online survey experiments show that a data marketplace increases participants' willingness to sell their X data by 12-25 percentage points compared to data donation (depending on treatments), and by 6.8 points compared to one-time purchase offers. Although difference in minimum acceptable prices are not statistically significant, over 64\% of participants set their price within the marketplace's suggested range (\$0.25-\$2)—substantially lower than the amounts offered in prior one-time purchase studies. Finally, in the marketplace setting, neither the type of buyer nor the inclusion of a privacy safeguard significantly influenced participants’ willingness to sell.

\end{abstract}

\newpage
\doublespacing

\section{Introduction}

Digital personal data are central to many critical issues in contemporary public discourse \cite{rathje2021out}. While users express privacy concerns over their data being used for AI training \cite{bengio2024managing}, and its potential harms \cite{hoes2025existential}, their sharing behavior contradicts these concerns—an inconsistency known as the \textit{`privacy paradox'} \cite{norberg2007privacy}. Meanwhile, researchers struggle to access platform data to study pressing societal issues such as mental health, misinformation, and content monetization in the post-API era \cite{bail2023we, alizadeh2023tokenization}. Existing regulations, such as the EU's AI Act, tend to prioritize risk reduction over public benefit \cite{prainsack2024new}. In response, researchers have explored `data donation' models \cite{araujo2022osd2f}, where individuals voluntarily share personal data for scientific purposes \cite{pffner2023leveraging}. Yet, these efforts face challenges in scalability due to ethical and legal constraints \cite{guerrini2018citizen}, high dropout, and low response rates \cite{strycharz2024blind}. Data marketplaces—where people sell their data to multiple buyers—have been proposed as a solution \cite{parra2020managers}, but their practical viability and the factors affecting user participation remain largely unexplored.

This paper examines social media users' willingness to sell their X (Twitter) data archives on a marketplace and their minimum acceptable price. It also explores how these decisions are affected by the type of buyer (researcher vs. private company) and the presence of a privacy-protection tool that automatically removes sensitive data before sharing. Social exchange theory \cite{thibaut2017social} suggests that willingness to engage is influenced by participants’ evaluation of potential rewards and costs. Building on this, Kmetty et al \cite{kmetty2024determinants} identified four key determinants of data-sharing behavior: incentives, task burden, data sensitivity, and participant characteristics including digital literacy, privacy concerns, and platform usage \cite{breuer2023user}. Of these, financial incentives have shown the strongest effect on data donation \cite{kmetty2024determinants}; however, their applicability to data marketplaces remains unexplored. Moreover, while previous studies typically offer \$5–\$10 for users' X data in one-time, non-marketplace offers \cite{kmetty2024determinants}, we explore whether a marketplace model can reduce this cost to under \$2. This benefits both parties—sellers can potentially earn more by accessing multiple buyers, while buyers pay significantly less per dataset.

Despite existing theoretical frameworks for data marketplaces \cite{parra2020managers}, little is known about their impact on individuals’ privacy concerns and willingness to sell their social media data. We address this gap with empirical evidence from two online survey experiments in the U.S. (total $N$ = 2,500). Study 1 examines how presence of a marketplace affects individuals' willingness to sell X data and their pricing, varying treatments by offer type—marketplace vs. single buyer—and suggested price range (\$0.25-\$1 and \$1-\$1.75). Study 2 examines the effects of buyer type (university researchers vs. private companies) and privacy protection. We also explore whether these effects are moderated by factors such as privacy concerns, privacy literacy, and demographics. We find that, depending on treatments, a data marketplace can increase participants' willingness to sell by 12-25 percentage points compared to data donation, and by 5-11 points compared to single offers by researchers. Although the differences in the reported minimum price are not statistically significant, we observe that over 64\% of participants were willing to sell their X data within the suggested price ranges. Finally, within the marketplace condition, neither the buyer type nor the inclusion of a privacy-protection feature significantly influenced willingness to sell. These results suggest that data marketplaces can address researchers’ data access challenges and companies’ transparency issues, while enabling users to benefit from their data.

\section{Results}

\subsection{Study 1}

To examine how a data marketplace affects individuals’ willingness to sell their X data to university researchers and their pricing, we conducted an online survey experiment with $N = 1,250$ U.S. participants. After reading a brief explanation about downloading their X data and the need for user-provided data in the post-API era, participants were asked about their willingness to sell their privacy-protected X data (a tool that removes all sensitive data before sharing) and their minimum price. All participants received a fixed reward, but treatments varied by: a) offer type (single offer vs. marketplace), and b) marketplace suggested price (see \emph{Materials and Methods}). To avoid a hypothetical bias, following \cite{benndorf2018willingness}, we used two Becker-DeGroot-Marschak (BDM) mechanisms to elicit users' willingness to sell (see \emph{Materials and Methods}). Overall, 67\% of the participants were willing to sell their data. Among those unwilling, the top concerns were identifying who would use the data (60\%), understanding the study’s purpose (58\%), and uncertainty about what their X data includes (45\%).

\begin{figure*}
\centering
\includegraphics[width=0.85\linewidth]{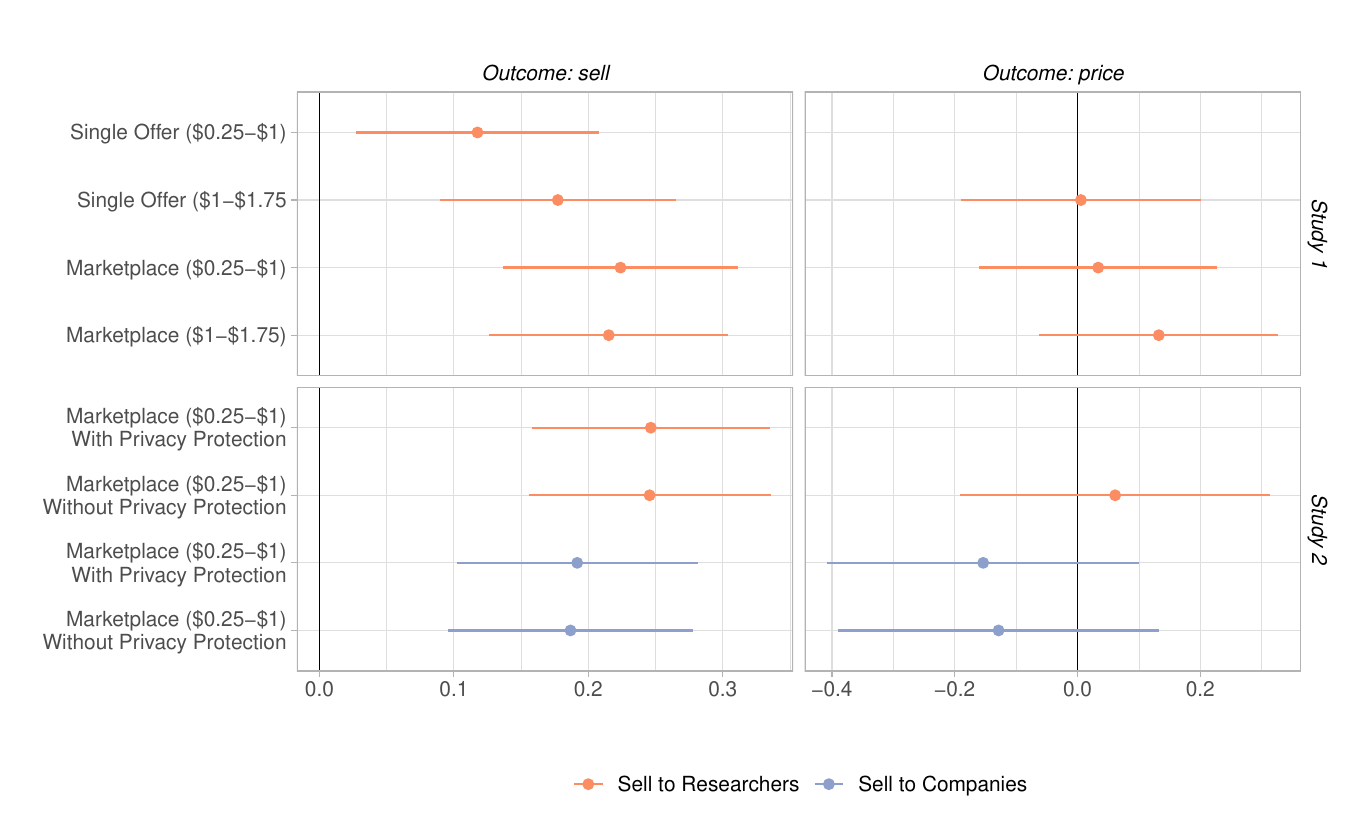}
\caption{Treatment effects from OLS regression with 95\% CI. Price effects are shown in standard deviations for comparability across Studies 1 and 2. A data marketplace increases willingness to sell by 12–25 percentage points, depending on the treatment. Selling to private companies reduces willingness to sell by 5-6 points, but the presence of a privacy protection feature has no effect. Treatments have no significant effect on requested price. In both studies, the control group is data donation (i.e. sharing for free).}
\label{fig:results}
\end{figure*}

The top panel of Figure \ref{fig:results} illustrates how offer type (single vs. marketplace) and suggested price range (\$0.25–\$1 vs. \$1–\$1.75) influence participants’ willingness to sell privacy-protected X data to university researchers across four experimental conditions: (1) a single researcher offering \$0.25–\$1; (2) a single researcher offering \$1–\$1.75; (3) a marketplace offer of \$0.25–\$1 requiring participation from at least 25 researchers; and (4) a similar marketplace offer priced at \$1–\$1.75.

Four key findings emerge. First, monetary incentives—regardless of whether offered individually or via a marketplace—increase willingness to sell by 12–25 percentage points relative to a data donation baseline. Second, neither treatment significantly influences the minimum price participants are willing to accept, although nearly 64\% of marketplace participants accepted prices within the suggested range—substantially lower than the $5–$10 range reported in prior studies. Third, increasing the suggested marketplace price does not yield a further increase in willingness to sell. Notably, the treatment effect was significantly moderated by participants’ privacy concerns (\textit{p} $<$ 0.05), whereas no other covariates had a statistically significant effect (see Materials and Methods for the list of covariates and Tables \ref{tab:reg_results_s1} \& \ref{tab:reg_results_min_s1} in the Appendix for regression table). Fourth, when pooling across treatment groups, participants in the marketplace condition are 6.8 percentage points more likely to agree to sell their X data compared to those in the single-offer condition (\textit{SE} = 0.032, \textit{p} = 0.035; Table \ref{tab:study1_pooled}).

\begin{table*}[!ht]
\centering
\renewcommand{\arraystretch}{1.3}
\caption{Effect of selling to researchers on marketplace compared to a single purchase offer.}
\footnotesize
\begin{tabular}{p{4cm}p{1.5cm}p{1.5cm}p{1.5cm}p{1.5cm}}
\hline
\textbf{Charactersitic} & \textbf{Estimate} & \textbf{Std. Err} & \textbf{t value} & \textbf{p\_value} \\
\hline
(Intercept) & -0.69 & 0.232 & -0.299 & 0.765 \\
marketplace & -0.68 & 0.032 & 2.110 & 0.035 * \\
as.factor(GENDER)2 & -0.098 & 0.034 & -2.856 & 0.004 ** \\
as.factor(GENDER)3  & 0.004 & 0.139 & 0.032 & 0.975 \\
as.factor(RACE)2  & -0.073 & 0.359 & -0.204 & 0.839 \\
as.factor(RACE)3  & 0.179 & 0.151 & 1.186 & 0.236 \\
\hline
$^{*}$p$<$0.05; $^{**}$p$<$0.01 & & & \\
\hline
\end{tabular}
\label{tab:study1_pooled}
\end{table*}

\subsection{Study 2}

To examine whether selling to private companies and removing sensitive data affects individuals’ willingness to sell their X data, we conducted a second experiment with 1,250 U.S. participants. After reading the same descriptions of their X data used in Study 1, participants reported their willingness to sell this data to small- and medium-sized private companies—specifically described as not being big tech firms. Treatments varied by: whether buyers were university researchers or private companies, and whether the marketplace remove sensitive data to ensure it never left users’ devices. We used a BDM mechanism with a minimum buyer-matching requirement to elicit users' willingness (see \emph{Materials and Methods}). Over 62\% of participants were willing to sell their data. Among those unwilling, top concerns included who would use the data (66\%), understanding the study’s purpose (64\%), and uncertainty about what their X data includes (40\%).

The bottom panel of Figure \ref{fig:results} presents the effects of buyer type (university researchers vs. private companies) and the availability of a sensitive data removal tool on participants’ willingness to sell their X data. Four key findings emerge. First, participants are, on average, 5.5 percentage points more likely to sell their data to researchers than to private companies across all conditions; however, this difference is not statistically significant (\textit{SE} = 0.032, \textit{p} = 0.088; see Table \ref{tab:study2_pooled}). Second, the presence of a privacy protection tool that automatically removes sensitive information prior to data sharing does not significantly influence willingness to sell. Third, while participants generally report lower asking prices when selling to private companies, this difference is also not statistically significant. Fourth, there is a descriptive pattern indicating a greater willingness to sell data to private companies on marketplaces (Study 2) compared to one-time offers to researchers (Study 1), though this comparison cannot be formally tested. Finally, no significant individual-level moderators were identified in the regression analysis (see Tables \ref{tab:reg_results_s2} \& \ref{tab:reg_results_min_s2} in Appendix).

\begin{table*}[!ht]
\centering
\renewcommand{\arraystretch}{1.3}
\caption{Effect of selling on marketplace compared to a single purchase offer}
\footnotesize
\begin{tabular}{p{6cm}p{1.5cm}p{1.5cm}p{1.5cm}p{1.5cm}}
\hline
\textbf{Charactersitic} & \textbf{Estimate} & \textbf{Std. Err} & \textbf{t value} & \textbf{p\_value} \\
\hline
(Intercept) & 0.371 & 0.269 & 1.379 & 0.168 \\
I(Group in c("T1", "T2"))TRUE & 0.055 & 0.032 & 1.70 & 0.089 \\
as.factor(GENDER)2 & -0.084 & 0.0338 & -2.494 & 0.013 \\
as.factor(GENDER)3  & 0.028 & 0.273 & 0.101 & 0.920 \\
as.factor(RACE)2  & -1.064 & 0.513 & -2.074 & 0.038 \\
as.factor(RACE)3  & -0.390 & 0.219 & -1.777 & 0.076 \\
\hline
$^{*}$p$<$0.05; $^{**}$p$<$0.01 & & & \\
\hline
\end{tabular}
\label{tab:study2_pooled}
\end{table*}

\section{Discussion}

We demonstrate that shifting from a traditional data donation model to a data marketplace has the potential to significantly increases individuals' willingness to sell their X (Twitter) data, even for modest compensation ranging from \$0.25 to \$2. This marketplace model enables more transparent and accessible data collection for researchers and companies, while simultaneously allowing individuals to retain ownership and benefit from their personal data. Notably, the inclusion of a privacy protection tool on the platform does not significantly influence willingness to sell—an observation consistent with the well-documented privacy paradox \cite{norberg2007privacy}, wherein stated privacy concerns do not align with actual data-sharing behavior.

These findings are particularly relevant in the context of growing concerns about the opacity of personal data usage in training AI systems, and in response to increasing calls from privacy researchers for more participatory and accountable approaches to AI development. Our results provide empirical \textit{proof-of-concept} support for long-standing proposals advocating personal data marketplaces as a viable and ethically sound mechanism for scalable data sharing.

While our primary goal was to provide a proof-of-concept for the idea of data marketplaces, the findings should be interpreted in light of several limitations. Most notably, questions remain about the external validity of the results \cite{findley2021external}, particularly in the context of repeated interactions between data sellers and buyers on amarketplace over time. Additionally, although we employed an incentivized mechanism to elicit participants’ willingness to sell their data, our conclusions require further validation through experiments conducted in an actual marketplace settings. Future research should explore key implementation challenges, such as the risk of data marketplaces generating new forms of data silos, the role of AI agents in supporting user participation (e.g., enhancing engagement and reducing dropout), and the privacy and security implications associated with deploying such agents \cite{alizadeh2025simple}.

\section{Materials and Methods}\label{section:matrials}
The study was approved by the Ethics Committee of the Faculty of Arts and Social Sciences, University of Zurich. The experimental design and data analysis plan were preregistered at aspredicted.com and are available to view for \hyperlink{https://aspredicted.org/m94m-phsc.pdf}{Study 1} and \hyperlink{https://aspredicted.org/wsjr-f72x.pdf}{Study 2}. Codes and data are available on a \hyperlink{https://osf.io/c6uj8/?view_only=89c5279cb89143c5afd3f289fafbcae7}{OSF} repository.   

\subsection{Sampling}
In both Study 1 and Study 2, we preregistered a target sample of 1,250 American participants on Prolific, using a pre-screener to include only those who had tweeted more than four times in the past year. In Study 1, 1,590 started the survey and 1,250 finished it. Only 24 participants reported having no social media accounts. Although 11\% of participants failed at least one manipulation check, we retained them in the dataset in accordance with our preregistration plan. The final sample (mean age = 35.50) consisted of 518 females, 727 males, and 5 other gender. In Study 2, 1,641 started the survey and 1,250 finished it. 34 participants reported having no social media accounts, and 8\% failed at least one manipulation check; these participants were also retained in the dataset as preregistered. The final sample (mean age = 35.02) included 523 females, 725 males, and 2 other gender.

\subsection{Elicitation of Willingness to Sell}
To avoid a hypothetical bias, we used two versions of the Becker-DeGroot-Marschak (BDM) mechanism (see \emph{Appendix SI}). In the single offer condition, participants were told that a random price—\$0.25–\$1 for Group 1 or \$1–\$1.75 for Group 2—had been generated by the computer, and the data exchange would occur only if their price was less than or equal to that value. In the marketplace condition, participants were told that buyers had already set their maximum offers. Here, the exchange would occur only if the participant's price fell below a randomly drawn value between \$0.25 and \$1, and if at least 25 potential buyers were willing to pay the participant’s stated price.

\subsection{Experimental Design}

At the outset, all participants were informed that those who completed the entire study would receive a flat payment of \$1.10 for their time. Consent was obtained prior to participation. In both studies, participants are presented with some general information about how they can access their X data package, and then randomly assigned to either the control group or one of four treatment groups. In Study 1, treatment conditions varied by the type of offer—either a marketplace or a single offer—and by price range (\$0.25–\$1 or \$1–\$1.75). Participants in the treatment groups were presented with statements about an online marketplace where they could sell their X data to university researchers. In Study 2, the two treatments varied by buyer type—university researchers vs. small- and medium-sized private companies—and by the presence or absence of a feature that automatically removes sensitive data. Participants in the second treatment were further told: "We created a tool that you can run it on your own PC or laptop, which automatically detects your sensitive data, removes them, and create a new ZIP file without those sensitive data. Then you can upload this new Zip file on our marketplace". Next, we used a BDM mechanism with a minimum buyer-matching requirement to elicit willing to sell for \$0.25-\$1, and if they answered `Yes', we asked them to indicate their minimum price. If they answered `No', we asked to select reason(s) from seven provided options (see \emph{Appendix SI}). In both studies, when the buyer was described as university researchers, we emphasized that all potential buyers had obtained ethics approval for their research. Afterward, all participants answered post-survey questions about privacy concerns, privacy literacy, and previous experience of privacy violation (see \emph{Appendix SI}). Finally, we debrief them about the goal of our study and that we will not actually ask them to share their X data with us (see \emph{Appendix SI}).

\subsection{Control Group}
In both Studies 1 and 2, the control condition followed the standard data donation procedure. Participants read a statement explaining how to download their X data and noting that, in the post-API era, researchers rely on user-provided data. The statement also assured them that all sensitive information would be removed before any data sharing. Participants were then asked: “Are you willing to donate your privacy-protected X (Twitter) data package to us for free under these conditions?”.

\subsection{Manipulation Check}
After each experiment, participants were asked: 1) Can you sell your data to more than one researcher, or just one researcher? (as many as available/just one), and 2) Can you choose to sell only parts of your X(Twitter) data, or do you have to sell the whole package? (I can choose/Only the whole package).

\subsection{Covariates}
The following covariates are measured in the pre- and post-survey sections: gender, age, race, party identification, social media accounts, social media use, privacy concerns (which we measure post-experimentally based on Westin's privacy index, as in Harris Interactive (2001)), privacy literacy including experience with privacy features, self-assessed privacy literacy, privacy rule application, and privacy rule adaptation, and previous experience of privacy violation. The full wordings are reported in \emph{Appendix SI}.

\section*{Acknowledgments}
\textbf{Funding:} This project received funding from the European Research Council (ERC) under the European Union's Horizon 2020 research and innovation program (grant agreement nr. 883121). \textbf{Author contribution:} MA conceived and designed the study. FG secured funding. Both authors contributed to data collection and analysis. MA drafted the initial manuscript. All authors revised the paper. We thank seminar participants at UZH Department of Political Science for their helpful feedback. Andy Guess, Ingmar Weber, Emma Hoes, and Sacha Altay shared ideas that we tried to incorporate. We thank Benjamin Streiff and Saba Yousefzadeh for outstanding research assistance. All errors are our own. \textbf{Competing interests:} The authors declare no competing interests. \textbf{Data and materials availability:} All data needed to evaluate the conclusions in the paper are present in the paper and/or the Supplementary Materials. 

\bibliographystyle{unsrt}

\newpage

\appendix
\pagenumbering{arabic}
\renewcommand{\thesection}{S\arabic{section}}

\section*{Full Results}
\subsection*{Study 1}

\begin{footnotesize}
    
\begin{longtable}[c]{llllll}
\caption{The effect of data marketplaces on willingness to share by all covariates in Study 1.}
\label{tab:reg_results_s1}\\
\hline
                            Characteristic  & Estimate  & Std. Error & t value & p\_value &    \\
\endfirsthead
\endhead
\hline
(Intercept)                        & -0.287382 & 0.359215   & -0.8    & 0.42389               &    \\
T1: Single Offer (\$0.25-\$1)          & 0.460285  & 0.519218   & 0.886   & 0.37557               &    \\
T2: Single Offer (\$1-\$2)             & -0.146488 & 0.465778   & -0.315  & 0.75321               &    \\
T3: Marketplace (\$0.25-\$1)           & 0.502887  & 0.530436   & 0.948   & 0.34333               &    \\
T4: Marketplace (\$1-\$2)              & 1.147975  & 0.494716   & 2.32    & 0.02052               & *  \\
Gender = Male                 & -0.147818 & 0.065312   & -2.263  & 0.02384               & *  \\
Gender = Female                 & -0.004202 & 0.352941   & -0.012  & 0.9905                &    \\
Race = Hawaiian                   & -0.557642 & 0.588696   & -0.947  & 0.34374               &    \\
Race = Asian                   & -0.484882 & 0.242726   & -1.998  & 0.04603               & *  \\
Race = African American                   & -0.325397 & 0.222021   & -1.466  & 0.14307               &    \\
Race = Hispanic                   & -0.403386 & 0.238385   & -1.692  & 0.09093               & .  \\
Race = White                   & -0.420896 & 0.217746   & -1.933  & 0.05353               & .  \\
Race = Other                   & -0.435574 & 0.269205   & -1.618  & 0.10598               &    \\
AGE                                & 0.034556  & 0.028879   & 1.197   & 0.23176               &    \\
EDUCATION                          & 0.004183  & 0.026416   & 0.158   & 0.87422               &    \\
IDEOLOGY                           & -0.043576 & 0.032898   & -1.325  & 0.18561               &    \\
Party = Republican         & 0.087085  & 0.086341   & 1.009   & 0.31341               &    \\
Having a social media account               & -0.015185 & 0.113691   & -0.134  & 0.89377               &    \\
Having Twitter/Reddit/etc         & 0.030872  & 0.017492   & 1.765   & 0.07788               & .  \\
Social Media Use                 & 0.026534  & 0.013206   & 2.009   & 0.04478               & *  \\
PrivacyConcerns                    & 0.234014  & 0.071204   & 3.287   & 0.00105               & ** \\
PrivacyRuleApplication             & -0.02554  & 0.079154   & -0.323  & 0.74702               &    \\
PrivacyViolationExperience         & 0.016823  & 0.043454   & 0.387   & 0.69873               &    \\
PrivacyLiteracy                    & 0.063608  & 0.02787    & 2.282   & 0.02269               & *  \\
PrivacyRuleAdaptation              & 0.006825  & 0.02479    & 0.275   & 0.78313               &    \\
GroupT1:Gender=Female         & -0.031881 & 0.097706   & -0.326  & 0.74427               &    \\
GroupT2:Gender=Female         & 0.0552    & 0.095657   & 0.577   & 0.56403               &    \\
GroupT3:Gender=Female         & 0.011996  & 0.094374   & 0.127   & 0.89888               &    \\
GroupT4:Gender=Female         & 0.202434  & 0.096337   & 2.101   & 0.03587               & *  \\
GroupT1:Gender=Other         & -0.219035 & 0.501292   & -0.437  & 0.66225               &    \\
GroupT2:Gender=Other         & 0.168546  & 0.494973   & 0.341   & 0.73354               &    \\
GroupT3:Gender=Other         & -0.10466  & 0.418222   & -0.25   & 0.80245               &    \\
GroupT4:Gender=Other         & 0.176763  & 0.452641   & 0.391   & 0.69624               &    \\
GroupT1:Race = Hawaiian           & NA        & NA         & NA      & NA                    &    \\
GroupT2:Race = Hawaiian           & 1.025149  & 0.798535   & 1.284   & 0.19952               &    \\
GroupT3:Race = Hawaiian           & NA        & NA         & NA      & NA                    &    \\
GroupT4:Race = Hawaiian           & NA        & NA         & NA      & NA                    &    \\
GroupT1:Race = Asian           & 0.538996  & 0.440079   & 1.225   & 0.22095               &    \\
GroupT2:Race = Asian           & 0.7582    & 0.38225    & 1.984   & 0.04759               & *  \\
GroupT3:Race = Asian           & 0.684259  & 0.431872   & 1.584   & 0.11342               &    \\
GroupT4:Race = Asian           & 0.572982  & 0.365593   & 1.567   & 0.11737               &    \\
GroupT1:Race = African American           & 0.320768  & 0.409865   & 0.783   & 0.43404               &    \\
GroupT2:Race = African American           & 0.556918  & 0.339667   & 1.64    & 0.10141               &    \\
GroupT3:Race = African American           & 0.36516   & 0.411393   & 0.888   & 0.37496               &    \\
GroupT4:Race = African American           & 0.224038  & 0.335626   & 0.668   & 0.5046                &    \\
GroupT1:Race = Hispanic           & 0.332439  & 0.428591   & 0.776   & 0.43814               &    \\
GroupT2:Race = Hispanic           & 0.760897  & 0.364057   & 2.09    & 0.03687               & *  \\
GroupT3:Race = Hispanic           & 0.49362   & 0.432307   & 1.142   & 0.2538                &    \\
GroupT4:Race = Hispanic           & 0.413535  & 0.362938   & 1.139   & 0.25481               &    \\
GroupT1:Race = White           & 0.523254  & 0.404262   & 1.294   & 0.19585               &    \\
GroupT2:Race = White           & 0.698081  & 0.331108   & 2.108   & 0.03526               & *  \\
GroupT3:Race = White           & 0.553382  & 0.404445   & 1.368   & 0.17155               &    \\
GroupT4:Race = White           & 0.191679  & 0.328242   & 0.584   & 0.55938               &    \\
GroupT1:Race = Other           & 0.165422  & 0.476462   & 0.347   & 0.72852               &    \\
GroupT2:Race = Other           & 0.875804  & 0.393896   & 2.223   & 0.02641               & *  \\
GroupT3:Race = Other           & 0.329821  & 0.460274   & 0.717   & 0.47381               &    \\
GroupT4:Race = Other           & 0.209219  & 0.402543   & 0.52    & 0.60336               &    \\
GroupT1:AGE                        & -0.034091 & 0.042068   & -0.81   & 0.41793               &    \\
GroupT2:AGE                        & -0.012362 & 0.040145   & -0.308  & 0.75821               &    \\
GroupT3:AGE                        & -0.043278 & 0.040769   & -1.062  & 0.2887                &    \\
GroupT4:AGE                        & -0.069094 & 0.042036   & -1.644  & 0.10056               &    \\
GroupT1:EDUCATION                  & 0.009483  & 0.039426   & 0.241   & 0.80997               &    \\
GroupT2:EDUCATION                  & -0.037325 & 0.037489   & -0.996  & 0.31969               &    \\
GroupT3:EDUCATION                  & 0.006987  & 0.038358   & 0.182   & 0.85551               &    \\
GroupT4:EDUCATION                  & -0.057096 & 0.039063   & -1.462  & 0.14416               &    \\
GroupT1:IDEOLOGY                   & 0.034883  & 0.048099   & 0.725   & 0.46848               &    \\
GroupT2:IDEOLOGY                   & 0.033721  & 0.049839   & 0.677   & 0.49883               &    \\
GroupT3:IDEOLOGY                   & 0.033138  & 0.049409   & 0.671   & 0.50257               &    \\
GroupT4:IDEOLOGY                   & 0.042499  & 0.048121   & 0.883   & 0.37736               &    \\
GroupT1:Party = Republican & -0.149403 & 0.132918   & -1.124  & 0.26128               &    \\
GroupT2:Party = Republican & 0.072893  & 0.128282   & 0.568   & 0.57001               &    \\
GroupT3:Party = Republican & -0.089534 & 0.13321    & -0.672  & 0.50166               &    \\
GroupT4:Party = Republican & -0.041304 & 0.128968   & -0.32   & 0.74884               &    \\
GroupT1:PrivacyConcerns            & -0.218977 & 0.100489   & -2.179  & 0.02956               & *  \\
GroupT2:PrivacyConcerns            & -0.190884 & 0.095779   & -1.993  & 0.04654               & *  \\
GroupT3:PrivacyConcerns            & -0.202323 & 0.102749   & -1.969  & 0.04922               & *  \\
GroupT4:PrivacyConcerns            & -0.274621 & 0.10661    & -2.576  & 0.01014               & *  \\
GroupT1:PrivacyRuleApplication     & -0.100388 & 0.126259   & -0.795  & 0.42675               &    \\
GroupT2:PrivacyRuleApplication     & 0.076757  & 0.119698   & 0.641   & 0.52151               &    \\
GroupT3:PrivacyRuleApplication     & -0.004873 & 0.118384   & -0.041  & 0.96717               &    \\
GroupT4:PrivacyRuleApplication     & 0.062907  & 0.124785   & 0.504   & 0.61429               &    \\
GroupT1:PrivacyViolationExperience & 0.061312  & 0.063906   & 0.959   & 0.3376                &    \\
GroupT2:PrivacyViolationExperience & 0.030368  & 0.062499   & 0.486   & 0.62715               &    \\
GroupT3:PrivacyViolationExperience & -0.037674 & 0.060832   & -0.619  & 0.53585               &    \\
GroupT4:PrivacyViolationExperience & 0.009802  & 0.062214   & 0.158   & 0.87485               &    \\
GroupT1:PrivacyLiteracy            & -0.00735  & 0.042033   & -0.175  & 0.86122               &    \\
GroupT2:PrivacyLiteracy            & 0.039237  & 0.039456   & 0.994   & 0.32025               &    \\
GroupT3:PrivacyLiteracy            & -0.02519  & 0.039146   & -0.643  & 0.52006               &    \\
GroupT4:PrivacyLiteracy            & -0.009544 & 0.042557   & -0.224  & 0.82259               &    \\
GroupT1:PrivacyRuleAdaptation      & -0.021308 & 0.036358   & -0.586  & 0.55797               &    \\
GroupT2:PrivacyRuleAdaptation      & 0.003524  & 0.034361   & 0.103   & 0.91834               &    \\
GroupT3:PrivacyRuleAdaptation      & -0.002812 & 0.037055   & -0.076  & 0.93952               &    \\
GroupT4:PrivacyRuleAdaptation      & -0.018548 & 0.037745   & -0.491  & 0.62325               &   \\
\hline
$^{.}$p$<$0.1; $^{*}$p$<$0.05; $^{**}$p$<$0.01 & & &\\
\hline
\end{longtable}

\end{footnotesize}

\newpage

\begin{footnotesize}
    
\begin{longtable}[c]{lllll}
\caption{The effect of data marketplaces on requested minimum price by all covariates in Study 1.}
\label{tab:reg_results_min_s1}\\
\hline
                               Characteristics & Estimate  & Std. Error & t value & p\_value \\
\endfirsthead
\endhead
\hline
(Intercept)                        & 106.186   & 571.6583   & 0.186   & 0.8527                \\
T2: Single Offer (\$1-\$2)         & -1.8267   & 742.3322   & -0.002  & 0.998                 \\
T3: Marketplace (\$0.25-\$1)       & 19.2756   & 735.262    & 0.026   & 0.9791                \\
T4: Marketplace (\$1-\$2)       & 832.7921  & 681.719    & 1.222   & 0.2225                \\
Gender = Female                 & -0.9171   & 98.3102    & -0.009  & 0.9926                \\
Gender = Other                 & 121.8573  & 469.5071   & 0.26    & 0.7953                \\
Race = Hawaiian                   & -64.2552  & 710.7148   & -0.09   & 0.928                 \\
Race = Asian                   & -62.7464  & 510.1621   & -0.123  & 0.9022                \\
Race = African American                   & -21.9856  & 483.8173   & -0.045  & 0.9638                \\
Race = Hispanic                   & -36.301   & 493.3394   & -0.074  & 0.9414                \\
Race = White                   & -51.4222  & 477.1449   & -0.108  & 0.9142                \\
Race = Other                   & -50.2743  & 583.954    & -0.086  & 0.9314                \\
AGE                                & -7.3936   & 40.1664    & -0.184  & 0.854                 \\
EDUCATION                          & 2.3334    & 39.5599    & 0.059   & 0.953                 \\
IDEOLOGY                           & 3.2071    & 47.7913    & 0.067   & 0.9465                \\
Party = Republican         & 22.6098   & 132.011    & 0.171   & 0.8641                \\
Having a social media account               & 38.8776   & 182.2415   & 0.213   & 0.8312                \\
Having Twitter/Reddit/etc               & 11.162    & 25.126     & 0.444   & 0.6571                \\
Social Media Use                 & -36.929   & 18.4578    & -2.001  & 0.0460 *              \\
PrivacyConcerns                    & -3.2678   & 91.6825    & -0.036  & 0.9716                \\
PrivacyRuleApplication             & 9.6184    & 121.6647   & 0.079   & 0.937                 \\
PrivacyViolationExperience         & -1.9354   & 57.4196    & -0.034  & 0.9731                \\
PrivacyLiteracy                    & -0.1513   & 40.1699    & -0.004  & 0.997                 \\
PrivacyRuleAdaptation              & 1.1595    & 34.7659    & 0.033   & 0.9734                \\
GroupT2:Gender=Female         & -10.671   & 130.5333   & -0.082  & 0.9349                \\
GroupT3:Gender=Female         & 43.0811   & 126.9496   & 0.339   & 0.7345                \\
GroupT4:Gender=Female         & -57.2987  & 128.8918   & -0.445  & 0.6568                \\
GroupT2:Gender=Other         & -182.8709 & 663.5814   & -0.276  & 0.783                 \\
GroupT3:Gender=Other         & -156.949  & 543.2552   & -0.289  & 0.7728                \\
GroupT4:Gender=Other         & -364.077  & 544.0716   & -0.669  & 0.5037                \\
GroupT2:Race = Hawaiian           & NA        & NA         & NA      & NA                    \\
GroupT3:Race = Hawaiian           & NA        & NA         & NA      & NA                    \\
GroupT4:Race = Hawaiian           & NA        & NA         & NA      & NA                    \\
GroupT2:Race = Asian           & -5.4137   & 769.1269   & -0.007  & 0.9944                \\
GroupT3:Race = Asian           & -4.4733   & 705.1346   & -0.006  & 0.9949                \\
GroupT4:Race = Asian           & -60.3248  & 592.0357   & -0.102  & 0.9189                \\
GroupT2:Race = African American           & -30.5308  & 709.6678   & -0.043  & 0.9657                \\
GroupT3:Race = African American           & 40.5288   & 679.901    & 0.06    & 0.9525                \\
GroupT4:Race = African American           & 165.9246  & 559.1797   & 0.297   & 0.7668                \\
GroupT2:Race = Hispanic           & -34.7632  & 732.8013   & -0.047  & 0.9622                \\
GroupT3:Race = Hispanic           & 4.0832    & 692.7586   & 0.006   & 0.9953                \\
GroupT4:Race = Hispanic           & -178.7559 & 576.912    & -0.31   & 0.7568                \\
GroupT2:Race = White           & -35.4277  & 706.7262   & -0.05   & 0.96                  \\
GroupT3:Race = White           & -1.1502   & 670.2144   & -0.002  & 0.9986                \\
GroupT4:Race = White           & -83.9239  & 550.2579   & -0.153  & 0.8788                \\
GroupT2:Race = Other           & -47.0293  & 795.8977   & -0.059  & 0.9529                \\
GroupT3:Race = Other           & -27.9772  & 782.1944   & -0.036  & 0.9715                \\
GroupT4:Race = Other           & -103.0601 & 684.0002   & -0.151  & 0.8803                \\
GroupT2:AGE                        & 9.4421    & 51.4755    & 0.183   & 0.8545                \\
GroupT3:AGE                        & 15.3384   & 52.9894    & 0.289   & 0.7724                \\
GroupT4:AGE                        & -14.7279  & 54.5164    & -0.27   & 0.7872                \\
GroupT2:EDUCATION                  & -1.8178   & 52.3022    & -0.035  & 0.9723                \\
GroupT3:EDUCATION                  & 5.4822    & 52.1325    & 0.105   & 0.9163                \\
GroupT4:EDUCATION                  & -29.8653  & 51.4284    & -0.581  & 0.5617                \\
GroupT2:IDEOLOGY                   & -0.4171   & 64.3963    & -0.006  & 0.9948                \\
GroupT3:IDEOLOGY                   & -10.6823  & 63.5232    & -0.168  & 0.8665                \\
GroupT4:IDEOLOGY                   & -75.9339  & 61.5691    & -1.233  & 0.2181                \\
GroupT2:Party = Republican & -55.5396  & 172.298    & -0.322  & 0.7473                \\
GroupT3:Party = Republican & -7.9787   & 178.546    & -0.045  & 0.9644                \\
GroupT4:Party = Republican & 319.7917  & 172.2342   & 1.857   & 0.0640 .              \\
GroupT2:PrivacyConcerns            & -1.2073   & 122.8396   & -0.01   & 0.9922                \\
GroupT3:PrivacyConcerns            & -32.8302  & 126.4599   & -0.26   & 0.7953                \\
GroupT4:PrivacyConcerns            & -236.1868 & 126.3472   & -1.869  & 0.0622 .              \\
GroupT2:PrivacyRuleApplication     & 20.6621   & 163.7716   & 0.126   & 0.8997                \\
GroupT3:PrivacyRuleApplication     & 40.8815   & 158.4424   & 0.258   & 0.7965                \\
GroupT4:PrivacyRuleApplication     & -48.0004  & 163.7144   & -0.293  & 0.7695                \\
GroupT2:PrivacyViolationExperience & 11.3476   & 79.9338    & 0.142   & 0.8872                \\
GroupT3:PrivacyViolationExperience & 24.4459   & 75.9326    & 0.322   & 0.7476                \\
GroupT4:PrivacyViolationExperience & 60.0746   & 77.3419    & 0.777   & 0.4377                \\
GroupT2:PrivacyLiteracy            & 4.0318    & 56.046     & 0.072   & 0.9427                \\
GroupT3:PrivacyLiteracy            & -15.2901  & 50.5893    & -0.302  & 0.7626                \\
GroupT4:PrivacyLiteracy            & 5.7283    & 55.2085    & 0.104   & 0.9174                \\
GroupT2:PrivacyRuleAdaptation      & 0.7237    & 44.2881    & 0.016   & 0.987                 \\
GroupT3:PrivacyRuleAdaptation      & 14.7521   & 47.1811    & 0.313   & 0.7547                \\
GroupT4:PrivacyRuleAdaptation      & 55.2617   & 47.3279    & 1.168   & 0.2435 \\             
\hline
$^{.}$p$<$0.1; $^{*}$p$<$0.05; $^{**}$p$<$0.01 & & &\\
\hline
\end{longtable}

\end{footnotesize}

\newpage

\subsection*{Study 2}
\begin{footnotesize}
    
\begin{longtable}[c]{llllll}
\caption{The effect of data marketplaces on willingness to share by all covariates in Study 2.}
\label{tab:reg_results_s2}\\
\hline

                                                                       Characteristic  & Estimate  & Std. Error & t value & p\_value &    \\
\endfirsthead
\endhead
\hline
(Intercept)                                                              & 0.916905  & 0.419143   & 2.188   & 0.02894               & *  \\
T1: Single Offer (\$0.25-\$1)          & -0.294442 & 0.689955   & -0.427  & 0.66965               &    \\
T2: Single Offer (\$1-\$2)                                                                  & -0.555186 & 0.674686   & -0.823  & 0.41078               &    \\
T3: Marketplace (\$0.25-\$1)              & -0.223667 & 0.598667   & -0.374  & 0.70878               &    \\
T4: Marketplace (\$1-\$2)               & -0.302315 & 0.697433   & -0.433  & 0.66477               &    \\
Gender = Female                                                       & 0.104704  & 0.066789   & 1.568   & 0.11728               &    \\
Gender = Other                                                       & -0.438563 & 0.479419   & -0.915  & 0.36053               &    \\
Race = Hawaiian                                                         & -1.093973 & 0.666156   & -1.642  & 0.10087               &    \\
Race = Asian                                                         & -0.773119 & 0.312181   & -2.477  & 0.01344               & *  \\
Race = African American                                                         & -0.569183 & 0.287859   & -1.977  & 0.04829               & *  \\
Race = Hispanic                                                         & -0.36377  & 0.308895   & -1.178  & 0.23922               &    \\
Race = White                                                         & -0.701362 & 0.283394   & -2.475  & 0.0135                & *  \\
Race = Other                                                         & -0.778079 & 0.330997   & -2.351  & 0.01894               & *  \\
AGE                                                                      & -0.021869 & 0.028457   & -0.768  & 0.44238               &    \\
EDUCATION                                                                & -0.019732 & 0.027714   & -0.712  & 0.47664               &    \\
IDEOLOGY                                                                 & 0.075162  & 0.033672   & 2.232   & 0.02583               & *  \\
Party = Republican                                               & -0.114224 & 0.088675   & -1.288  & 0.19801               &    \\
Having a social media account                                                     & -0.065916 & 0.089288   & -0.738  & 0.46055               &    \\
Having Twitter/Reddit/etc                                                     & 0.016472  & 0.017078   & 0.965   & 0.33501               &    \\
Social Media Use                                                       & 0.012956  & 0.013006   & 0.996   & 0.31943               &    \\
PrivacyConcerns                                                          & 0.021727  & 0.071752   & 0.303   & 0.7621                &    \\
PrivacyRuleApplication                                                   & -0.127951 & 0.091635   & -1.396  & 0.16294               &    \\
PrivacyViolationExperience                                               & -0.087802 & 0.048665   & -1.804  & 0.0715                & .  \\
PrivacyLiteracy                                                          & 0.005093  & 0.029592   & 0.172   & 0.86338               &    \\
PrivacyRuleAdaptation                                                    & 0.039187  & 0.025442   & 1.54    & 0.12382               &    \\
GroupT1:Gender=Female                                               & -0.145838 & 0.096447   & -1.512  & 0.13083               &    \\
GroupT2:Gender=Female                                               & -0.160753 & 0.09651    & -1.666  & 0.0961                & .  \\
GroupT3:Gender=Female                                               & -0.228615 & 0.096114   & -2.379  & 0.01757               & *  \\
GroupT4:Gender=Female                                               & -0.242514 & 0.095599   & -2.537  & 0.01134               & *  \\
GroupT1:Gender=Other                                               & 0.749525  & 0.588195   & 1.274   & 0.20287               &    \\
GroupT2:Gender=Other                                               & NA        & NA         & NA      & NA                    &    \\
GroupT3:Gender=Other                                               & NA        & NA         & NA      & NA                    &    \\
GroupT4:Gender=Other                                               & NA        & NA         & NA      & NA                    &    \\
GroupT1:Race = Hawaiian                                                 & NA        & NA         & NA      & NA                    &    \\
GroupT2:Race = Hawaiian                                                 & NA        & NA         & NA      & NA                    &    \\
GroupT3:Race = Hawaiian                                                 & NA        & NA         & NA      & NA                    &    \\
GroupT4:Race = Hawaiian                                                 & NA        & NA         & NA      & NA                    &    \\
GroupT1:Race = Asian                                                 & 0.39788   & 0.581576   & 0.684   & 0.49405               &    \\
GroupT2:Race = Asian                                                 & 0.488928  & 0.591333   & 0.827   & 0.40854               &    \\
GroupT3:Race = Asian                                                 & 0.210294  & 0.471729   & 0.446   & 0.65585               &    \\
GroupT4:Race = Asian                                                 & 0.566578  & 0.58641    & 0.966   & 0.33419               &    \\
GroupT1:Race = African American                                                 & 0.308955  & 0.557257   & 0.554   & 0.57942               &    \\
GroupT2:Race = African American                                                 & 0.219525  & 0.564093   & 0.389   & 0.69724               &    \\
GroupT3:Race = African American                                                 & 0.125901  & 0.44437    & 0.283   & 0.77699               &    \\
GroupT4:Race = African American                                                 & 0.171031  & 0.558447   & 0.306   & 0.75947               &    \\
GroupT1:Race = Hispanic                                                 & -0.185468 & 0.582714   & -0.318  & 0.75034               &    \\
GroupT2:Race = Hispanic                                                 & 0.228152  & 0.590201   & 0.387   & 0.69916               &    \\
GroupT3:Race = Hispanic                                                 & -0.421337 & 0.510962   & -0.825  & 0.4098                &    \\
GroupT4:Race = Hispanic                                                 & 0.087359  & 0.590777   & 0.148   & 0.88248               &    \\
GroupT1:Race = White                                                 & 0.379253  & 0.555468   & 0.683   & 0.49492               &    \\
GroupT2:Race = White                                                 & 0.52994   & 0.559601   & 0.947   & 0.34388               &    \\
GroupT3:Race = White                                                 & 0.254954  & 0.444162   & 0.574   & 0.56609               &    \\
GroupT4:Race = White                                                 & 0.441401  & 0.558675   & 0.79    & 0.42967               &    \\
GroupT1:Race = Other                                                 & 0.470646  & 0.600803   & 0.783   & 0.43361               &    \\
GroupT2:Race = Other                                                 & 0.742901  & 0.657478   & 1.13    & 0.25879               &    \\
GroupT3:Race = Other                                                 & -0.024719 & 0.526483   & -0.047  & 0.96256               &    \\
GroupT4:Race = Other                                                 & 0.325121  & 0.611139   & 0.532   & 0.59485               &    \\
GroupT1:AGE                                                              & 0.043949  & 0.040302   & 1.09    & 0.27577               &    \\
GroupT2:AGE                                                              & -0.050113 & 0.040641   & -1.233  & 0.21785               &    \\
GroupT3:AGE                                                              & 0.07071   & 0.039689   & 1.782   & 0.07512               & .  \\
GroupT4:AGE                                                              & -0.043363 & 0.042176   & -1.028  & 0.30414               &    \\
GroupT1:EDUCATION                                                        & 0.016732  & 0.040213   & 0.416   & 0.67744               &    \\
GroupT2:EDUCATION                                                        & 0.042452  & 0.040234   & 1.055   & 0.29163               &    \\
GroupT3:EDUCATION                                                        & -0.021407 & 0.039782   & -0.538  & 0.59064               &    \\
GroupT4:EDUCATION                                                        & 0.015871  & 0.041143   & 0.386   & 0.69976               &    \\
GroupT1:IDEOLOGY                                                         & -0.126703 & 0.047375   & -2.674  & 0.00761               & ** \\
GroupT2:IDEOLOGY                                                         & -0.01095  & 0.046589   & -0.235  & 0.81422               &    \\
GroupT3:IDEOLOGY                                                         & -0.077589 & 0.044186   & -1.756  & 0.07941               & .  \\
GroupT4:IDEOLOGY                                                         & -0.080708 & 0.047689   & -1.692  & 0.0909                & .  \\
GroupT1:Party = Republican                                       & 0.112709  & 0.125587   & 0.897   & 0.3697                &    \\
GroupT2:Party = Republican                                       & 0.005658  & 0.124329   & 0.046   & 0.96371               &    \\
GroupT3:Party = Republican                                       & 0.11988   & 0.118748   & 1.01    & 0.31297               &    \\
GroupT4:Party = Republican                                       & 0.178396  & 0.127261   & 1.402   & 0.16129               &    \\
GroupT1:PrivacyConcerns                                                  & 0.031013  & 0.102339   & 0.303   & 0.76192               &    \\
GroupT2:PrivacyConcerns                                                  & -0.004924 & 0.103777   & -0.047  & 0.96216               &    \\
GroupT3:PrivacyConcerns                                                  & -0.007153 & 0.103686   & -0.069  & 0.94501               &    \\
GroupT4:PrivacyConcerns                                                  & 0.124338  & 0.098866   & 1.258   & 0.20882               &    \\
GroupT1:PrivacyRuleApplication                                           & 0.205306  & 0.12423    & 1.653   & 0.09873               & .  \\
GroupT2:PrivacyRuleApplication                                           & 0.04872   & 0.129749   & 0.375   & 0.70738               &    \\
GroupT3:PrivacyRuleApplication                                           & 0.071588  & 0.122414   & 0.585   & 0.55882               &    \\
GroupT4:PrivacyRuleApplication                                           & 0.016637  & 0.128421   & 0.13    & 0.89695               &    \\
GroupT1:PrivacyViolationExperience                                       & 0.041943  & 0.066946   & 0.627   & 0.53112               &    \\
GroupT2:PrivacyViolationExperience                                       & 0.052263  & 0.067163   & 0.778   & 0.43667               &    \\
GroupT3:PrivacyViolationExperience                                       & 0.108967  & 0.066284   & 1.644   & 0.10051               &    \\
GroupT4:PrivacyViolationExperience                                       & 0.046562  & 0.066555   & 0.7     & 0.48434               &    \\
GroupT1:PrivacyLiteracy                                                  & 0.033495  & 0.0424     & 0.79    & 0.42974               &    \\
GroupT2:PrivacyLiteracy                                                  & 0.094624  & 0.044687   & 2.118   & 0.03447               & *  \\
GroupT3:PrivacyLiteracy                                                  & 0.069982  & 0.042286   & 1.655   & 0.09825               & .  \\
GroupT4:PrivacyLiteracy                                                  & 0.037844  & 0.040646   & 0.931   & 0.35205               &    \\
GroupT1:PrivacyRuleAdaptation                                            & -0.014079 & 0.036173   & -0.389  & 0.69722               &    \\
GroupT2:PrivacyRuleAdaptation                                            & -0.02732  & 0.036445   & -0.75   & 0.45367               &    \\
GroupT3:PrivacyRuleAdaptation                                            & -0.015561 & 0.035486   & -0.439  & 0.66112               &    \\
GroupT4:PrivacyRuleAdaptation                                            & -0.0278   & 0.036084   & -0.77   & 0.44124               &    \\
GroupT4:PrivacyRuleAdaptation      & -0.027800   &0.036084 & -0.770 & 0.44124   \\
\hline
$^{.}$p$<$0.1; $^{*}$p$<$0.05; $^{**}$p$<$0.01 & & &\\
\hline
\end{longtable}
\end{footnotesize}

\newpage

\begin{footnotesize}
    
\begin{longtable}[c]{lllll}
\caption{The effect of data marketplaces on requested minimum price by all covariates in Study 2.}
\label{tab:reg_results_min_s2}\\
\hline
                            Characteristic       & Estimate  & Std. Error & t value & Pr(\textgreater{}|t|) \\
\endfirsthead
\endhead
\hline
(Intercept)                        & -0.66292  & 55.41979   & -0.012  & 0.990461              \\
T2: Single Offer (\$1-\$2)                            & -20.37379 & 75.17068   & -0.271  & 0.786484              \\
T3: Marketplace (\$0.25-\$1)                            & 1.22664   & 69.78678   & 0.018   & 0.985984              \\
T4: Marketplace (\$1-\$2)                            & -1.30039  & 77.75281   & -0.017  & 0.986663              \\
Gender = Female                 & 11.6932   & 7.96543    & 1.468   & 0.142772              \\
Gender = Other                 & 7.98421   & 31.12218   & 0.257   & 0.797643              \\
Race = Asian                   & 4.87413   & 45.80261   & 0.106   & 0.915298              \\
Race = African American                   & 11.6178   & 43.48432   & 0.267   & 0.789453              \\
Race = Hispanic                   & 4.47954   & 47.32296   & 0.095   & 0.924626              \\
Race = White                   & 5.00055   & 43.60953   & 0.115   & 0.908758              \\
Race = Other                   & -3.35014  & 46.85308   & -0.072  & 0.943028              \\
AGE                                & -0.58246  & 3.25655    & -0.179  & 0.858126              \\
EDUCATION                          & -1.84807  & 3.2781     & -0.564  & 0.573183              \\
IDEOLOGY                           & 4.08931   & 3.67174    & 1.114   & 0.265963              \\
Party = Republican         & 1.26609   & 9.97542    & 0.127   & 0.899057              \\
Having a social media account               & 7.03809   & 11.87706   & 0.593   & 0.553747              \\
Having Twitter/Reddit/etc               & 0.03201   & 2.25416    & 0.014   & 0.988675              \\
Social Media Use                 & -1.43649  & 1.63814    & -0.877  & 0.380985              \\
PrivacyConcerns                    & -2.16038  & 8.76704    & -0.246  & 0.805464              \\
PrivacyRuleApplication             & -0.71269  & 9.36959    & -0.076  & 0.939401              \\
PrivacyViolationExperience         & -3.88655  & 5.04403    & -0.771  & 0.441374              \\
PrivacyLiteracy                    & -0.07063  & 3.73731    & -0.019  & 0.98493               \\
PrivacyRuleAdaptation              & 1.00995   & 2.74951    & 0.367   & 0.713545              \\
GroupT2:Gender=Female         & 6.61285   & 11.16122   & 0.592   & 0.55381               \\
GroupT3:Gender=Female         & -11.42344 & 11.67208   & -0.979  & 0.328231              \\
GroupT4:Gender=Female         & -11.84599 & 11.56736   & -1.024  & 0.306319              \\
GroupT2:Gender=Other         & NA        & NA         & NA      & NA                    \\
GroupT3:Gender=Other         & NA        & NA         & NA      & NA                    \\
GroupT4:Gender=Other         & NA        & NA         & NA      & NA                    \\
GroupT2:Race = Asian           & 14.91573  & 65.87279   & 0.226   & 0.820963              \\
GroupT3:Race = Asian           & -5.57808  & 57.13284   & -0.098  & 0.922265              \\
GroupT4:Race = Asian           & -6.18106  & 65.27609   & -0.095  & 0.9246                \\
GroupT2:Race = African American           & 27.2149   & 62.42261   & 0.436   & 0.663052              \\
GroupT3:Race = African American           & -8.84325  & 53.40659   & -0.166  & 0.868556              \\
GroupT4:Race = African American           & -13.20087 & 61.82395   & -0.214  & 0.831011              \\
GroupT2:Race = Hispanic           & 11.21024  & 66.65979   & 0.168   & 0.866521              \\
GroupT3:Race = Hispanic           & -12.20523 & 65.24416   & -0.187  & 0.851686              \\
GroupT4:Race = Hispanic           & -3.78336  & 67.11345   & -0.056  & 0.955069              \\
GroupT2:Race = White           & -0.3702   & 62.16327   & -0.006  & 0.995251              \\
GroupT3:Race = White           & -5.13559  & 53.83555   & -0.095  & 0.924042              \\
GroupT4:Race = White           & -6.49492  & 62.12201   & -0.105  & 0.916777              \\
GroupT2:Race = Other           & 36.18034  & 70.39286   & 0.514   & 0.607508              \\
GroupT3:Race = Other           & 4.74975   & 71.29484   & 0.067   & 0.946911              \\
GroupT4:Race = Other           & 2.00257   & 69.15444   & 0.029   & 0.97691               \\
GroupT2:AGE                        & 17.15518  & 4.71444    & 3.639   & 0.000304 ***          \\
GroupT3:AGE                        & -0.091    & 4.44773    & -0.02   & 0.983686              \\
GroupT4:AGE                        & 0.72631   & 5.05645    & 0.144   & 0.885845              \\
GroupT2:EDUCATION                  & -1.38179  & 4.68961    & -0.295  & 0.768391              \\
GroupT3:EDUCATION                  & 1.17482   & 4.52437    & 0.26    & 0.795235              \\
GroupT4:EDUCATION                  & 1.29101   & 4.90274    & 0.263   & 0.792416              \\
GroupT2:IDEOLOGY                   & 0.49711   & 5.20756    & 0.095   & 0.923991              \\
GroupT3:IDEOLOGY                   & -3.90088  & 4.86044    & -0.803  & 0.422623              \\
GroupT4:IDEOLOGY                   & -4.23922  & 5.37264    & -0.789  & 0.430485              \\
GroupT2:Party = Republican & 15.62227  & 13.90601   & 1.123   & 0.261831              \\
GroupT3:Party = Republican & -3.33803  & 13.2103    & -0.253  & 0.800622              \\
GroupT4:Party = Republican & -1.44939  & 14.96215   & -0.097  & 0.92287               \\
GroupT2:PrivacyConcerns            & -2.36919  & 11.98802   & -0.198  & 0.84342               \\
GroupT3:PrivacyConcerns            & 3.75836   & 12.56974   & 0.299   & 0.765071              \\
GroupT4:PrivacyConcerns            & 4.12537   & 12.50979   & 0.33    & 0.741719              \\
GroupT2:PrivacyRuleApplication     & -1.59965  & 13.75274   & -0.116  & 0.907452              \\
GroupT3:PrivacyRuleApplication     & -0.14679  & 13.20205   & -0.011  & 0.991133              \\
GroupT4:PrivacyRuleApplication     & -0.23153  & 14.22      & -0.016  & 0.987016              \\
GroupT2:PrivacyViolationExperience & -3.57269  & 7.47564    & -0.478  & 0.632935              \\
GroupT3:PrivacyViolationExperience & 3.73894   & 7.41705    & 0.504   & 0.614425              \\
GroupT4:PrivacyViolationExperience & 4.0531    & 7.625      & 0.532   & 0.595284              \\
GroupT2:PrivacyLiteracy            & -10.52794 & 5.3928     & -1.952  & 0.051503 .            \\
GroupT3:PrivacyLiteracy            & 0.8192    & 5.31823    & 0.154   & 0.877646              \\
GroupT4:PrivacyLiteracy            & -0.1338   & 4.92876    & -0.027  & 0.978354              \\
GroupT2:PrivacyRuleAdaptation      & 3.18152   & 3.88932    & 0.818   & 0.413762              \\
GroupT3:PrivacyRuleAdaptation      & -1.92346  & 3.9393     & -0.488  & 0.625583              \\
GroupT4:PrivacyRuleAdaptation      & -0.32874  & 3.95919    & -0.083  & 0.933862 \\   
\hline
$^{.}$p$<$0.1; $^{*}$p$<$0.05; $^{**}$p$<$0.01 & & &\\
\hline
\end{longtable}

\end{footnotesize}

\newpage

\section*{Survey Items Study 1}

\subsection*{Study 1 -- Informed Consent}
\vspace{1ex}
\noindent
\setstretch{1.2}

Please read and accept the terms and conditions within 1 minute if you wish to continue, else your participation will be terminated. Do not refresh the page at any point during the survey. At the end of the survey you will receive a completion code. 

\textbf{Attention:} To participate in this survey, you must have an active X(Twitter) account. Otherwise you won't receive compensation. We appreciate your interest in participating in this study. You have been invited to take part as you have registered for this task on Prolific. Please note that you may only participate in this study if you are 18 years of age or over.

Please read through the information provided and accept the consent form below, if you wish to proceed. The aim of this study is to investigate people’s willingness to share their X(Twitter) data with university researchers. As a participant, you will be invited to read some information about a scenario we have in mind. Some of the information we provide may not reflect reality.

Before and after joining the study, you will be asked basic questions about your age, demographics, social media use, and your opinions on topics such as privacy concerns, privacy literacy, and privacy violations. This should take about 5 minutes. No background knowledge is required.

\textbf{Do I have to take part?} Participation is voluntary. You may withdraw at any point before submitting your answers by pressing the ‘Exit’ button or closing the browser. Note: We only remunerate participants who complete the entire study.

\textbf{Will I be compensated?} Yes. Participants with an active X(Twitter) account who complete the study will be paid a flat rate of \$1.10.

\textbf{How will your data be used?} Your answers will be completely anonymous and stored securely. Data may be used in academic publications, but your IP address will not be recorded. Data will be stored for a minimum of three years after publication.

\textbf{By selecting "I consent" below you confirm that you:}

\hspace*{1em}have read the above information\\
\hspace*{1em}voluntarily agree to participate\\
\hspace*{1em}are at least 18 years of age\\
\hspace*{1em}have a genuine active X(Twitter) account

\begin{itemize}
    \item I consent
    \item I do not consent
\end{itemize}

\subsection*{Study 1 -- Pre-Survey}
\vspace{1ex}
\noindent
\setstretch{1.2}

\textbf{GENDER}\\
How do you describe yourself?
\begin{enumerate}
    \item Male
    \item Female
    \item Other
\end{enumerate}

\textbf{AGE}\\
Which of the following categories includes your age?
\begin{enumerate}
    \item 17 or younger
    \item 18 to 24
    \item 25 to 34
    \item 35 to 44
    \item 45 to 54
    \item 55 to 64
    \item 65+
\end{enumerate}

\textbf{RACE}\\
How do you describe yourself? (Please check the one option that best describes you)
\begin{enumerate}
    \item American Indian or Alaskan Native
    \item Hawaiian or other Pacific islander
    \item Asian or Asian American
    \item Black or African American
    \item Hispanic or Latino
    \item Non-Hispanic White
    \item Other
\end{enumerate}

\textbf{EDUCATION}\\
What is the highest level of education you have completed?
\begin{enumerate}
    \item None, or grades 1-8
    \item High school incomplete (grades 9-11)
    \item High school graduate (grade 12 or GED certificate)
    \item Technical, trade or vocational school AFTER high school
    \item Some college, no 4-year degree (includes associate degree)
    \item College graduate (B.S., B.A., or other 4-year degree)
    \item Post-graduate training/professional school after college (toward a Master's degree or Ph.D., Law or Medical school)
\end{enumerate}

\textbf{IDEOLOGY}\\
In politics, people sometimes talk about liberal and conservative. In general, would you describe yourself as:
\begin{enumerate}
    \item Very liberal
    \item Somewhat liberal
    \item Moderate
    \item Somewhat conservative
    \item Very conservative
\end{enumerate}

\textbf{PID -- FORCED}\\
If you absolutely had to choose between only the Democratic and Republican Party, which do you prefer?
\begin{enumerate}
    \item Democratic Party
    \item Republican Party
\end{enumerate}

\textbf{SOCIAL MEDIA ACC 1}\\
Do you have a social media profile (e.g., Facebook, X(Twitter), Instagram, Reddit, TikTok)?
\begin{enumerate}
    \item No
    \item Yes
\end{enumerate}

\textbf{SOCIAL MEDIA ACC 2}\\
Do you have accounts on any of the following social media sites? Please select all that apply
\begin{itemize}
    \item X(Twitter)
    \item Instagram
    \item Reddit
    \item TikTok
    \item None
\end{itemize}

\textbf{SOCIAL MEDIA USE}\\
On a typical day, how much time would you say you spend on social media (such as Facebook and X(Twitter); either on a mobile or a computer
\begin{enumerate}
    \item Less than 30 minutes
    \item 30-60 minutes
    \item 1-2 hours
    \item 2-3 hours
    \item 3+ hours
\end{enumerate}

\subsection*{Study 1 -- Treatment 1 -- Intro}
\vspace{1ex}
\noindent
\setstretch{1.2}

\textbf{S1 T1 INTRO}\\
Social media platforms like X(Twitter) now allow users to export and download their data as a ZIP file. This means all X(Twitter) users can get a full copy of their data. Since X(Twitter) has closed its data access API, researchers now rely on individuals who are willing to share their data. We’ve created a tool at our university where anyone interested in selling their privacy-protected X(Twitter) data package to us for our university research can sign up, set a minimum price, and (if that price is accepted) sell it to us. You can run the tool on your own PC or laptop. Once you upload your X (Twitter) data on that tool, it automatically detects your sensitive data (e.g. direct messages, phone number, emails of persons in your contact list), removes them, and creates a new privacy-protected ZIP file (i.e. without those sensitive data). \textbf{Our research project received ethics approval from our university’s ethics committee.}

\textbf{Here’s how it works:} The computer has randomly generated an amount of money \textbf{between \$0.25 and \$1} to offer you to sell your privacy-protected package of X(Twitter) data to us for research. Before we tell you what the offer is, we will ask you the smallest offer you would be willing to accept. If the offer the computer generated is above the amount you give, we will send you simple step-by-step instructions to help you download your X(Twitter) data, run it through our tool to remove sensitive data, send it to us through a secure channel, and pay you the offered amount if you do. If the offer is below the amount you give, we will not ask you to sell (you won’t have another chance to adjust your price). \textbf{Regardless of the price match, all participants who complete the study will receive a flat rate of \$1.10 as a thank you for their time.}

\textbf{Example 1:} Computer generated \$0.8. You set a minimum price of \$1, so you don’t pass the eligibility step. Then, we will pay you \$1.10 for participating in this study after you complete it. But won’t ask you to sell your data to us. 

\textbf{Example 2:} Computer generated \$0.8. You set a minimum price of \$0.6, so you pass the eligibility step. Then, we will pay you \$1.10 for participating in this study. We would later ask you to send us your data and if you do, we pay another \$0.6 for your data.

\subsection*{Study 1 -- Manipulation check}
\vspace{1ex}
\noindent
\setstretch{1.2}
\hspace*{1em}\textit{Displayed after each treatment INTRO}

\textbf{MANIPULATION CHECK PRIMER}\\
Please answer the following questions based on the scenario you have just read:

\textbf{MANIPULATION CHECK Q1}\\
Can you sell your data to more than one researcher, or just one researcher?
\begin{itemize}
    \item As many researchers as available
    \item Just one researcher
\end{itemize}

\textbf{MANIPULATION CHECK Q2}\\
Can you choose to sell only parts of your X(Twitter) data, or do you have to sell the whole data package?
\begin{itemize}
    \item I could choose the data types that I'm willing to sell
    \item I could only sell the complete privacy-protected package
\end{itemize}

\subsection*{Study 1 -- Treatment 1}
\vspace{1ex}
\noindent
\setstretch{1.2}

\textbf{SELL? S1 T1}\\
Are you \textbf{willing to sell us} your privacy-protected X(Twitter) data package under these conditions?
\begin{itemize}
    \item No
    \item Yes
\end{itemize}

\textbf{MINPRICE S1 T1}\\ 
What is \textbf{your minimum price} in two-decimal \$USD format? (e.g. 0.45, 1.00)\\
\hspace*{1em}\textit{Displayed as open textbox if "Yes" was selected in SELL? S1 T1}

\textbf{WILLING S1 T1}\\
In addition to receiving money, is there any other motivation for you in selling your data to researchers? (optional)
\hspace*{1em} \textit{Displayed as open textbox if "Yes" was selected in SELL? S1 T1}

\textbf{NOTWILLING S1 T1}\\
Can you tell us why you don’t want to sell your data to researchers?\\
\hspace*{1em} \textit{Displayed if "No" was selected in SELL? S1 T1}

\begin{enumerate}
    \item I need to know exactly who will be using my data
    \item I need to understand the specific purpose for which my data is needed
    \item I don't trust X(Twitter)
    \item I don't trust researchers
    \item I trust researchers but am unsure whether they share data with third parties
    \item I am unsure about what is included in my X(Twitter) data
    \item Other
\end{enumerate}

\textbf{S1 T1 OTHER}\\
You have selected "other". Please specify:\\
\hspace*{1em} \textit{Displayed as open textbox if "Other" was selected in NOTWILLING S1 T1}


\subsection*{Study 1 -- Treatment 2 -- Intro}
\vspace{1ex}
\noindent
\setstretch{1.2}

\textbf{S1 T2 INTRO}\\
Social media platforms like X(Twitter) now allow users to export and download their data as a ZIP file. This means all X(Twitter) users can get a full copy of their data. Since X(Twitter) has closed its data access API, researchers now rely on individuals who are willing to share their data. We’ve created a tool at our university where anyone interested in selling their privacy-protected X(Twitter) data package to us for our university research can sign up, set a minimum price, and (if that price is accepted) sell it to us. You can run the tool on your own PC or laptop. Once you upload your X (Twitter) data on that tool, it automatically detects your sensitive data (e.g. direct messages, phone number, emails of persons in your contact list), removes them, and creates a new privacy-protected ZIP file (i.e. without those sensitive data). \textbf{Our research project received ethics approval from our university’s ethics committee.}

\textbf{Here’s how it works:} The computer has randomly generated an amount of money \textbf{between \$1 and \$1.75} to offer you to sell your privacy-protected package of X(Twitter) data to us for research. Before we tell you what the offer is, we will ask you the smallest offer you would be willing to accept. If the offer the computer generated is above the amount you give, we will send you simple step-by-step instructions to help you download your X(Twitter) data, run it through our tool to remove sensitive data, send it to us through a secure channel, and pay you the offered amount if you do. If the offer is below the amount you give, we will not ask you to sell (you won’t have another chance to adjust your price). \textbf{Regardless of the price match, all participants who complete the study will receive a flat rate of \$1.10 as a thank you for their time.}

\textbf{Example 1:} Computer generated \$1.4. You set a minimum price of \$1.5, so you don’t pass the eligibility step. Then, we will pay you \$1.10 for participating in this study after you complete it. But won’t ask you to sell your data to us.

\textbf{Example 2:} Computer generated \$1.4. You set a minimum price of \$1.2, so you pass the eligibility step. Then, we will pay you \$1.10 for participating in this study after you complete it. We would later ask you to send us your data and if you do, we pay another \$1.2 for your data.

\subsection*{Study 1 -- Treatment 2}
\vspace{1ex}
\noindent
\setstretch{1.2}

\textbf{SELL? S1 T2}\\
Are you \textbf{willing to sell us} your privacy-protected X(Twitter) data package under these conditions?
\begin{itemize}
    \item No
    \item Yes
\end{itemize}

\textbf{MINPRICE S1 T2}\\ 
What is \textbf{your minimum price} in two-decimal \$USD format? (e.g. 1.00, 1.35)\\
\hspace*{1em}\textit{Displayed as open textbox if "Yes" was selected in SELL? S1 T2}

\textbf{WILLING S1 T2}\\
In addition to receiving money, is there any other motivation for you in selling your data to researchers? (optional)\\
\hspace*{1em}\textit{Displayed as open textbox if "Yes" was selected in SELL? S1 T2}

\textbf{NOTWILLING S1 T2}\\
Can you tell us why you don’t want to sell your data to researchers?\\
\hspace*{1em}\textit{Displayed if "No" was selected in SELL? S1 T2}
\begin{enumerate}
    \item I need to know exactly who will be using my data
    \item I need to understand the specific purpose for which my data is needed
    \item I don't trust X(Twitter)
    \item I don't trust researchers
    \item I trust researchers but am unsure whether they share data with third parties
    \item I am unsure about what is included in my X(Twitter) data
    \item Other
\end{enumerate}

\textbf{S1 T2 OTHER}\\
You have selected "other". Please specify:\\
\hspace*{1em}\textit{Displayed as open textbox if "Other" was selected in NOTWILLING S1 T2}


\subsection*{Study 1 -- Treatment 3 -- Intro}
\vspace{1ex}
\noindent
\setstretch{1.2}

\textbf{S1 T3 INTRO}\\
Social media platforms like X(Twitter) now allow users to export and download their data as a ZIP file. This means all X(Twitter) users can get a full copy of their data. Since X(Twitter) has closed its data access API, researchers now rely on individuals who are willing to share their data. We’ve created a marketplace at our university for X(Twitter) data where anyone interested in selling their privacy-protected X(Twitter) data package to university researchers can sign up, set a minimum price, and sell it to \textbf{as many university researchers as possible.} In other words, you can sell the same data to many researchers just by uploading it on our marketplace. We also created a tool that you can run it on your own PC or laptop, which automatically detects your sensitive data (e.g. direct messages, phone number, emails of persons in your contact list), removes them, and create a new privacy-protected ZIP file (i.e. without those sensitive data). Then you can upload this new ZIP file on our marketplace. Since launching two weeks ago, we’ve already received offers from buyers who have set their maximum prices. \textbf{All our buyers are university researchers who have received ethics approval from their institutions’ ethics committees.}

\textbf{Here’s how it works:} Based on the maximum prices researchers on our marketplace are willing to pay, the computer has randomly generated an amount of money \textbf{between \$0.25 and \$1} to offer you to sell your privacy-protected package of X(Twitter) data to matched researchers. Before we tell you what the offer is, we will ask you the smallest offer you would be willing to accept. If the offer the computer generated is above the amount you give \textbf{(Condition 1)}, AND at least 25 buyers match your price \textbf{(Condition 2)}, we will send you simple step-by-step instructions to help you download your X (Twitter) data, run it through our tool to remove sensitive data, upload it on our marketplace, and pay you the total amount from all matching buyers. If the offer is below that amount, OR fewer than 25 buyers matched your price, we will not ask you to sell (you won’t have another chance to adjust your price). \textbf{Regardless of the match, all participants who complete the study will receive a flat rate of \$1.10 as a thank you for their time.}

\textbf{Example 1:} Computer generated \$0.8. You set a minimum price of \$1, so you don’t pass Condition 1. Then, we will pay you \$1.10 for participating in this study after you complete it. The marketplace would pay you nothing because you failed Condition 1.

\textbf{Example 2:} Computer generated \$0.8. You set a minimum price of \$0.6, so you pass Condition 1. Your price is equal or lower than the maximum price of only 20 buyers, which is less than the required minimum of 25 buyers, so you do NOT pass Condition 2. Then, we will pay you \$1.10 for participating in this study after you complete it. The marketplace would pay you nothing because you failed Condition 2.

\textbf{Example 3:} Computer generated \$0.8. You set a minimum price of \$0.6, so you pass Condition 1. Your price is equal or lower than the maximum price of 30 buyers, so you pass Condition 2 as well. Then we will pay you \$1.10 for participating in this study after you complete it. The marketplace would pay you \$0.6 × 30 = \$18.

\subsection*{Study 1 -- Treatment 3}
\vspace{1ex}
\noindent
\setstretch{1.2}

\textbf{SELL? S1 T3}\\
Are you \textbf{willing to sell} your privacy-protected X(Twitter) data package under these conditions?
\begin{itemize}
    \item No
    \item Yes
\end{itemize}

\textbf{MINPRICE S1 T3}\\
What is \textbf{your minimum price} in two-decimal \$USD format? (e.g. 0.45, 1.00)\\
\hspace*{1em}\textit{Displayed as open textbox if "Yes" was selected in SELL? S1 T3}

\textbf{WILLING S1 T3}\\
In addition to receiving money, is there any other motivation for you in selling your data to researchers? (optional)\\
\hspace*{1em}\textit{Displayed as open textbox if "Yes" was selected in SELL? S1 T3}

\textbf{NOTWILLING S1 T3}\\
Can you tell us why you don’t want to sell your data to researchers?\\
\hspace*{1em}\textit{Displayed if "No" was selected in SELL? S1 T3}
\begin{enumerate}
    \item I need to know exactly who will be using my data
    \item I need to understand the specific purpose for which my data is needed
    \item I don't trust X(Twitter)
    \item I don't trust researchers
    \item I trust researchers but am unsure whether they share data with third parties
    \item I am unsure about what is included in my X(Twitter) data
    \item Other
\end{enumerate}

\textbf{S1 T3 OTHER}\\
You have selected "other". Please specify:\\
\hspace*{1em}\textit{Displayed as open textbox if "Other" was selected in NOTWILLING S1 T3}


\subsection*{Study 1 -- Treatment 4 -- Intro}
\vspace{1ex}
\noindent
\setstretch{1.2}

\textbf{S1 T4 INTRO}\\
Social media platforms like X(Twitter) now allow users to export and download their data as a ZIP file. This means all X(Twitter) users can get a full copy of their data. Since X(Twitter) has closed its data access API, researchers now rely on individuals who are willing to share their data. We’ve created a marketplace at our university for X(Twitter) data where anyone interested in selling their privacy-protected X(Twitter) data package to university researchers can sign up, set a minimum price, and sell it to \textbf{as many university researchers as possible.} In other words, you can sell the same data to many researchers just by uploading it on our marketplace. We also created a tool that you can run it on your own PC or laptop, which automatically detects your sensitive data (e.g. direct messages, phone number, emails of persons in your contact list), removes them, and create a new privacy-protected ZIP file (i.e. without those sensitive data). Then you can upload this new ZIP file on our marketplace. Since launching two weeks ago, we’ve already received offers from buyers who have set their maximum prices. \textbf{All our buyers are university researchers who have received ethics approval from their institutions’ ethics committees.}

\textbf{Here’s how it works:} Based on the maximum prices researchers on our marketplace are willing to pay, the computer has randomly generated an amount of money \textbf{between \$1 and \$1.75} to offer you to sell your privacy-protected package of X(Twitter) data to matched researchers. Before we tell you what the offer is, we will ask you the smallest offer you would be willing to accept. If the offer the computer generated is above the amount you give \textbf{(Condition 1)}, AND at least 25 buyers match your price \textbf{(Condition 2)}, we will send you simple step-by-step instructions to help you download your X (Twitter) data, run it through our tool to remove sensitive data, upload it on our marketplace, and pay you the total amount from all matching buyers. If the offer is below that amount, OR fewer than 25 buyers matched your price, we will not ask you to sell (you won’t have another chance to adjust your price). \textbf{Regardless of the match, all participants who complete the study will receive a flat rate of \$1.10 as a thank you for their time.}

\textbf{Example 1:} Computer generated \$1.4. You set a minimum price of \$1.5, so you don’t pass condition 1. Then, we will pay you \$1.10 for participating in this study after you complete it. The marketplace would pay you nothing because you failed Condition 1.

\textbf{Example 2:} Computer generated \$1.4. You set a minimum price of \$1.2, so you pass condition 1. Your price is equal or lower than the maximum price of only 20 buyers, which is less than the required minimum of 25 buyers, so you do NOT pass condition 2. Then, we will pay you \$1.10 for participating in this study after you complete it. The marketplace would pay you nothing because you failed Condition 2.

\textbf{Example 3:} Computer generated \$1.4. You set a minimum price of \$1.2, so you pass Condition 1. Your price is equal or lower than the maximum price of 30 buyers, so you pass Condition 2 as well. Then we will pay you \$1.10 for participating in this study after you complete it. The marketplace would pay you \$1.2 × 30 = \$36.

\subsection*{Study 1 -- Treatment 4}
\vspace{1ex}
\noindent
\setstretch{1.2}

\textbf{SELL? S1 T4}\\
Are you \textbf{willing to sell} your privacy-protected X(Twitter) data package under these conditions?
\begin{itemize}
    \item No
    \item Yes
\end{itemize}

\textbf{MINPRICE S1 T4}\\
What is \textbf{your minimum price} in two-decimal \$USD format? (e.g. 1.00, 1.35)\\
\hspace*{1em}\textit{Displayed as open textbox if "Yes" was selected in SELL? S1 T4}

\textbf{WILLING S1 T4}\\
In addition to receiving money, is there any other motivation for you in selling your data to researchers? (optional)\\
\hspace*{1em}\textit{Displayed as open textbox if "Yes" was selected in SELL? S1 T4}

\textbf{NOTWILLING S1 T4}\\
Can you tell us why you don’t want to sell your data to researchers?\\
\hspace*{1em}\textit{Displayed if "No" was selected in SELL? S1 T4}
\begin{enumerate}
    \item I need to know exactly who will be using my data
    \item I need to understand the specific purpose for which my data is needed
    \item I don't trust X(Twitter)
    \item I don't trust researchers
    \item I trust researchers but am unsure whether they share data with third parties
    \item I am unsure about what is included in my X(Twitter) data
    \item Other
\end{enumerate}

\textbf{S1 T4 OTHER}\\
You have selected "other". Please specify:\\
\hspace*{1em}\textit{Displayed as open textbox if "Other" was selected in NOTWILLING S1 T4}


\subsection*{Study 1 -- Control -- Intro}
\vspace{1ex}
\noindent
\setstretch{1.2}

\textbf{S1 CONTROL INTRO}\\
Social media platforms like X(Twitter) now allow users to export and download their data as a ZIP file. This means all X(Twitter) users can get a full copy of their data. Since X(Twitter) has closed its data access API, researchers now rely on individuals who are willing to share their data. We’ve created a tool at our university where anyone interested in donating their privacy-protected X(Twitter) data package to us for research can sign up. You can run the tool on your PC or laptop. Once you upload your X(Twitter) data on that tool, it automatically detects your sensitive data (e.g. direct messages, phone number, emails of persons in your contact list), removes them, and creates a new privacy-protected ZIP file (i.e. without those sensitive data). Then you can send us this new privacy-protected ZIP file through a secure channel. We will send you simple step-by-step instructions to help you download your X(Twitter) data. Regardless of whether you donate your data or not, all participants who complete the study will receive a flat rate of \$1.10 as a thank you for their time. \textbf{Our research project received ethics approval from our university’s ethics committee.}

\subsection*{Study 1 -- Control}
\vspace{1ex}
\noindent
\setstretch{1.2}

\textbf{DONATE? S1 CONTROL}\\
Are you \textbf{willing to donate} your privacy-protected X(Twitter) data package to us \textbf{for free} under these conditions?
\begin{itemize}
    \item No
    \item Yes
\end{itemize}

\textbf{WILLING S1 CONTROL}\\
Can you tell us about your motivation for donating your data to university researchers? (optional)\\
\hspace*{1em}\textit{Displayed as open textbox if "Yes" was selected in DONATE? S1 CONTROL}

\textbf{NOTWILL S1 CONTROL}\\
Can you tell us why you don’t want to donate your data to researchers?\\
\hspace*{1em}\textit{Displayed if "No" was selected in DONATE? S1 CONTROL}
\begin{enumerate}
    \item I need to know exactly who will be using my data
    \item I need to understand the specific purpose for which my data is needed
    \item I don't trust X(Twitter)
    \item I don't trust researchers
    \item I trust researchers but am unsure whether they share data with third parties
    \item I am unsure about what is included in my X(Twitter) data
    \item Other
\end{enumerate}

\textbf{S1 CONTROL OTHER}\\
You have selected "other". Please specify:\\
\hspace*{1em}\textit{Displayed as open textbox if "Other" was selected in NOTWILL S1 CONTROL}


\subsection*{Study 1 -- Post-Survey}
\vspace{1ex}
\noindent
\setstretch{1.2}

\textbf{POST-SURVEY-INTRO}\\
In the following questions, we are curious to find out your knowledge on online data privacy and security.

\textbf{S2.1}\\
Please indicate below to which degree you agree/disagree with each of the following statements.\\
\hspace*{1em}\textit{Scale: 1 = Strongly disagree, 2 = Somewhat disagree, 3 = Somewhat agree, 4 = Strongly agree}

\begin{itemize}
    \item Consumers have lost all control over how personal information is collected and used by companies.
    \item Most businesses handle the personal information they collect about consumers in a proper and confidential way.
    \item Existing laws and organizational practices provide a reasonable level of protection for consumer privacy today.
\end{itemize}

\textbf{S2.2.1-1}\\
Have you ever blocked anyone on your X(Twitter) account?
\begin{itemize}
    \item Yes
    \item No
\end{itemize}

\textbf{S2.2.1-2}\\
Have you ever muted anyone on your X(Twitter) account?
\begin{itemize}
    \item Yes
    \item No
\end{itemize}

\textbf{S2.2.1-3}\\
Have you ever reported anyone on your X(Twitter) account?
\begin{itemize}
    \item Yes
    \item No
\end{itemize}

\textbf{S2.2.1-4}\\
Have you experienced previous privacy violations?
\begin{itemize}
    \item No
    \item Yes
    \item Yes, more than once
    \item I don't know
\end{itemize}

\textbf{S2.2.2}\\
Please indicate below to which degree you agree/disagree with each of the following statements.\\
\hspace*{1em}\textit{Scale: 1 = Strongly disagree, 2 = Disagree, 3 = Somewhat disagree, 4 = Neither agree nor disagree, 5 = Somewhat agree, 6 = Agree, 7 = Strongly agree}

\begin{itemize}
    \item I have the knowledge necessary to use privacy features to regulate information on X(Twitter).
    \item Given the knowledge it takes to use the privacy features, it would be easy for me to control information flow on X(Twitter).
\end{itemize}

\textbf{S2.2.3 and S2.2.4}\\
How frequently do you use privacy features on X(Twitter)?\\
\hspace*{1em}\textit{Scale: 1 = Never, 2 = Rarely, 3 = Occasionally, 4 = Sometimes, 5 = Frequently, 6 = Usually, 7 = Always}

\begin{itemize}
    \item How frequently do you post to the selected group of friends using privacy features (e.g., Twitter lists)?
    \item How frequently do you post a status update excluding some friends?
    \item How frequently have you deleted posts that others made on your timeline?
    \item How frequently have you untagged yourself in a photo or post that was posted by others?
\end{itemize}

\subsection*{Study 1 -- Debriefing}
\vspace{1ex}
\noindent
\setstretch{1.2}
Thank you for your participation in this research study. Now that you completed or have ended your participation, we will provide you with some additional information about the purposes of this study.
 
\textbf{What you should know about this study:}
The main goal of this study was to see how likely people are to sell all their X(Twitter) data and to understand the minimum price they would set for it. We haven’t actually created a marketplace yet, so we won’t be asking you to download or send your X(Twitter) data. We mentioned having such a platform to help create a realistic experience for participants
 
\textbf{If you have questions:}
The main researchers conducting this study are \textit{(blinded for peer review)}. If you have questions, you may contact one of the main researchers. If you have any questions or concerns regarding your rights as a research participant in this study, you may contact the Ethics Committee:
\textit{(contact info Ethics Committee)}


\section*{Survey Items Study 2}

\subsection*{Study 2 -- Informed Consent}
\vspace{1ex}
\noindent
\setstretch{1.2}

Please read and accept the terms and conditions within 1 minute if you wish to continue, else your participation will be terminated. Do not refresh the page at any point during the survey. At the end of the survey you will be redirected to Prolific.

\textbf{Attention:} To participate in this survey, you must have an active X(Twitter) account. Otherwise you won't receive compensation. We appreciate your interest in participating in this study. You have been invited to take part as you have registered for this task on Prolific. Please note that you may only participate in this study if you are 18 years of age or over.

Please read through the information provided and accept the consent form below, if you wish to proceed. The aim of this study is to investigate people’s willingness to share their X(Twitter) data with university researchers. As a participant, you will be invited to read some information about a scenario we have in mind. For the purpose of our study, some of the information we provide may not reflect the reality.

Before and after joining the study, you will be asked some basic questions about your age, demographics, social media use, and your opinion on a number of topics including privacy concerns, privacy literacy, and experience of privacy violations. This should take about 5 minutes. No background knowledge is required.

\textbf{Do I have to take part?} Participation is voluntary. You may withdraw at any point during the questionnaire before submitting your answers, by pressing the ‘Exit’ button or closing the browser. Note: We only remunerate participants who complete the entire study.

\textbf{Will I be compensated?} Yes. Participants who have an active X(Twitter) account and complete the study will be paid a flat rate of \$1.10.

\textbf{How will your data be used?} Your answers will be completely anonymous, and we will take reasonable steps to keep them confidential. Your data will be stored in a password-protected file and may be used in academic publications. Your IP address will not be stored. Research data will be retained for a minimum of three years after publication or public release.

\textbf{By selecting "I consent" below you confirm that you:}

\hspace*{1em}have read the above information\\
\hspace*{1em}voluntarily agree to participate\\
\hspace*{1em}are at least 18 years of age\\
\hspace*{1em}have a genuine active X(Twitter) account

\begin{itemize}
    \item I consent
    \item I do not consent
\end{itemize}

\subsection*{Study 2 -- Pre-Survey}
\vspace{1ex}
\noindent
\setstretch{1.2}

\textbf{GENDER}\\
How do you describe yourself?
\begin{enumerate}
    \item Male
    \item Female
    \item Other
\end{enumerate}

\textbf{AGE}\\
Which of the following categories includes your age?
\begin{enumerate}
    \item 17 or younger
    \item 18 to 24
    \item 25 to 34
    \item 35 to 44
    \item 45 to 54
    \item 55 to 64
    \item 65+
\end{enumerate}

\textbf{RACE}\\
How do you describe yourself? (Please check the one option that best describes you)
\begin{enumerate}
    \item American Indian or Alaskan Native
    \item Hawaiian or other Pacific Islander
    \item Asian or Asian American
    \item Black or African American
    \item Hispanic or Latino
    \item Non-Hispanic White
    \item Other
\end{enumerate}

\textbf{EDUCATION}\\
What is the highest level of education you have completed?
\begin{enumerate}
    \item None, or grades 1-8
    \item High school incomplete (grades 9-11)
    \item High school graduate (grade 12 or GED certificate)
    \item Technical, trade or vocational school AFTER high school
    \item Some college, no 4-year degree (includes associate degree)
    \item College graduate (B.S., B.A., or other 4-year degree)
    \item Post-graduate training/professional school after college (toward a Master's degree or Ph.D., Law or Medical school)
\end{enumerate}

\textbf{IDEOLOGY}\\
In politics, people sometimes talk about liberal and conservative. In general, would you describe yourself as:
\begin{enumerate}
    \item Very liberal
    \item Somewhat liberal
    \item Moderate
    \item Somewhat conservative
    \item Very conservative
\end{enumerate}

\textbf{PID -- FORCED}\\
If you absolutely had to choose between only the Democratic and Republican Party, which do you prefer?
\begin{enumerate}
    \item Democratic Party
    \item Republican Party
\end{enumerate}

\textbf{SOCIAL MEDIA ACC 1}\\
Do you have a social media profile (e.g., Facebook, X(Twitter), Instagram, Reddit, TikTok)?
\begin{enumerate}
    \item No
    \item Yes
\end{enumerate}

\textbf{SOCIAL MEDIA ACC 2}\\
Do you have accounts on any of the following social media sites? Please select all that apply.
\begin{itemize}
    \item X(Twitter)
    \item Instagram
    \item Reddit
    \item TikTok
    \item None
\end{itemize}

\textbf{SOCIAL MEDIA USE}\\
On a typical day, how much time would you say you spend on social media (such as Facebook and X(Twitter); either on a mobile or a computer)?
\begin{enumerate}
    \item Less than 30 minutes
    \item 30–60 minutes
    \item 1–2 hours
    \item 2–3 hours
    \item 3+ hours
\end{enumerate}


\subsection*{Study 2 -- Treatment 1 -- Intro}
\vspace{1ex}
\noindent
\setstretch{1.2}

\textbf{S2 T1 INTRO}\\
Social media platforms like X(Twitter) now allow users to export and download their data as a ZIP file. This means all X(Twitter) users can get a full copy of their data. Since X(Twitter) has closed its data access API, researchers now rely on individuals who are willing to share their data. 

We’ve created a marketplace for X(Twitter) data where anyone interested in selling their X(Twitter) data package to university researchers can sign up, set a minimum price, and sell it to \textbf{as many university researchers as possible.} We also created a tool that you can run on your own PC or laptop, which \textbf{automatically detects your sensitive data (e.g. direct messages, phone number)}, removes them, and creates a new ZIP file without those sensitive data. Then you can upload this new ZIP file on our marketplace. Since launching two weeks ago, we’ve already received offers from buyers who have set their maximum prices. \textbf{All our buyers are university researchers.}

\textbf{Here’s how it works:} Based on the maximum prices researchers on our marketplace are willing to pay, the computer has randomly generated an amount of money \textbf{between \$0.25 and \$1} to offer you to sell your sensitive-data-removed package of X(Twitter) data to matched researchers. Before we tell you what the offer is, we will ask you the smallest offer you would be willing to accept. If the offer the computer generated is above the amount you give \textbf{(Condition 1)}, AND at least 25 buyers match your price \textbf{(Condition 2)}, we will ask you to download your X(Twitter) data, run it through our tool to remove sensitive data, upload it on our marketplace, and pay you the total amount from all matching buyers. If the offer is below that amount, OR fewer than 25 buyers matched your price, we will not ask you to sell (you won’t have another chance to adjust your price). \textbf{Regardless of the match, all participants who complete the study will receive a flat rate of \$1.10 as a thank you for their time.}

\textbf{Example 1:} Computer generated \$0.8. You set a minimum price of \$1, so you don’t pass Condition 1. Then, we will pay you \$1.10 for participating in this study. The marketplace would pay you nothing because you failed Condition 1.

\textbf{Example 2:} Computer generated \$0.8. You set a minimum price of \$0.6, so you pass Condition 1. Your price is equal or lower than the maximum price of only 20 buyers, which is less than the required minimum of 25 buyers, so you do NOT pass condition 2. Then, we will pay you \$1.10 for participating in this study. The marketplace would pay you nothing because you failed Condition 2.

\textbf{Example 3:} Computer generated \$0.8. You set a minimum price of \$0.6, so you pass Condition 1. Your price is equal or lower than the maximum price of 40 buyers, so you pass Condition 2 as well. Then we will pay you \$1.10 for participating in this study. The marketplace would pay you \$0.6 × 40 = \$24.

\subsection*{Study 2 -- Treatment 1}
\vspace{1ex}
\noindent
\setstretch{1.2}

\textbf{SELL? S2 T1}\\
Are you willing to sell your X(Twitter) data package under these conditions?
\begin{itemize}
    \item Yes
    \item No
\end{itemize}

\textbf{MINPRICE S2 T1}\\
What is \textbf{your minimum price} in two-decimal \$USD format? (e.g. 0.25, 1.00)\\
\hspace*{1em}\textit{Displayed as open textbox if "Yes" was selected in SELL? S2 T1}

\textbf{WILLING S2 T1}\\
In addition to receiving money, is there any other motivation for you in selling your data? (optional)\\
\hspace*{1em}\textit{Displayed as open textbox if "Yes" was selected in SELL? S2 T1}

\textbf{NOTWILLING S2 T1}\\
Can you tell us why you don’t want to sell your data?
\begin{enumerate}
    \item I need to know exactly who will be using my data
    \item I need to understand the specific purpose for which my data is needed
    \item I don't trust X(Twitter)
    \item I don't trust researchers
    \item I trust researchers but am unsure whether they share data with third parties
    \item I am unsure about what is included in my X(Twitter) data
    \item Other
\end{enumerate}

\textbf{S2 T1 OTHER}\\
You have selected "other". Please specify:\\
\hspace*{1em}\textit{Displayed as open textbox if "Other" was selected in NOTWILLING S2 T1}


\subsection*{Study 2 -- Treatment 2 -- Intro}
\vspace{1ex}
\noindent
\setstretch{1.2}

\textbf{S2 T2 INTRO}\\
Social media platforms like X(Twitter) now allow users to export and download their data as a ZIP file. This means all X(Twitter) users can get a full copy of their data. Since X(Twitter) has closed its data access API, researchers now rely on individuals who are willing to share their data.

We’ve created a marketplace for X(Twitter) data where anyone interested in selling their X(Twitter) data package to university researchers can sign up, set a minimum price, and sell it to \textbf{as many university researchers as possible.} Since launching two weeks ago, we’ve already received offers from buyers who have set their maximum prices. \textbf{All our buyers are university researchers.}

\textbf{Here’s how it works:} Based on the maximum prices researchers on our marketplace are willing to pay, the computer has randomly generated an amount of money \textbf{between \$0.25 and \$1} to offer you to sell your package of X(Twitter) data to matched researchers. Before we tell you what the offer is, we will ask you the smallest offer you would be willing to accept. If the offer the computer generated is above the amount you give \textbf{(Condition 1)}, AND at least 25 buyers match your price \textbf{(Condition 2)}, we will ask you to download your X(Twitter) data, upload it on our marketplace, and pay you the total amount from all matching buyers. If the offer is below that amount, OR fewer than 25 buyers matched your price, we will not ask you to sell (you won’t have another chance to adjust your price). \textbf{Regardless of the match, all participants who complete the study will receive a flat rate of \$1.10 as a thank you for their time.}

\textbf{Example 1:} Computer generated \$0.8. You set a minimum price of \$1, so you don’t pass Condition 1. Then, we will pay you \$1.10 for participating in this study. The marketplace would pay you nothing because you failed Condition 1.

\textbf{Example 2:} Computer generated \$0.8. You set a minimum price of \$0.6, so you pass Condition 1. Your price is equal or lower than the maximum price of only 20 buyers, which is less than the required minimum of 25 buyers, so you do NOT pass condition 2. Then, we will pay you \$1.10 for participating in this study. The marketplace would pay you nothing because you failed Condition 2.

\textbf{Example 3:} Computer generated \$0.8. You set a minimum price of \$0.6, so you pass Condition 1. Your price is equal or lower than the maximum price of 40 buyers, so you pass Condition 2 as well. Then we will pay you \$1.10 for participating in this study. The marketplace would pay you \$0.6 × 40 = \$24.

\subsection*{Study 2 -- Treatment 2}
\vspace{1ex}
\noindent
\setstretch{1.2}

\textbf{SELL? S2 T2}\\
Are you willing to sell your X(Twitter) data package under these conditions?
\begin{itemize}
    \item Yes
    \item No
\end{itemize}

\textbf{MINPRICE S2 T2}\\
What is \textbf{your minimum price} in two-decimal \$USD format? (e.g. 0.25, 1.00)\\
\hspace*{1em}\textit{Displayed as open textbox if "Yes" was selected in SELL? S2 T2}

\textbf{WILLING S2 T2}\\
In addition to receiving money, is there any other motivation for you in selling your data? (optional)\\
\hspace*{1em}\textit{Displayed as open textbox if "Yes" was selected in SELL? S2 T2}

\textbf{NOTWILLING S2 T2}\\
Can you tell us why you don’t want to sell your data?
\begin{enumerate}
    \item I need to know exactly who will be using my data
    \item I need to understand the specific purpose for which my data is needed
    \item I don't trust X(Twitter)
    \item I don't trust researchers
    \item I trust researchers but am unsure whether they share data with third parties
    \item I am unsure about what is included in my X(Twitter) data
    \item Other
\end{enumerate}

\textbf{S2 T2 OTHER}\\
You have selected "other". Please specify:\\
\hspace*{1em}\textit{Displayed as open textbox if "Other" was selected in NOTWILLING S2 T2}


\subsection*{Study 2 -- Treatment 3 -- Intro}
\vspace{1ex}
\noindent
\setstretch{1.2}

\textbf{S2 T3 INTRO}\\
Social media platforms like X(Twitter) now allow users to export and download their data as a ZIP file. This means all X(Twitter) users can get a full copy of their data. Since X(Twitter) has closed its data access API, small and medium private companies now rely on individuals who are willing to share their data.

We’ve created a marketplace for X(Twitter) data where anyone interested in selling their X(Twitter) data package to small and medium private companies can sign up, set a minimum price, and sell it to \textbf{as many small and medium private companies as possible.} We also created a tool that you can run on your own PC or laptop, which automatically detects your sensitive data (e.g. direct messages, phone number), removes them, and creates a new ZIP file without those sensitive data. Then you can upload this new ZIP file on our marketplace. Since launching two weeks ago, we’ve already received offers from buyers who have set their maximum prices. \textbf{All our buyers are small and medium size private companies. We do NOT allow big tech companies like Google and OpenAI to use our marketplace.}

\textbf{Here’s how it works:} Based on the maximum prices the private companies on our marketplace are willing to pay, the computer has randomly generated an amount of money \textbf{between \$0.25 and \$1} to offer you to sell your sensitive-data-removed package of X(Twitter) data to matched companies. Before we tell you what the offer is, we will ask you the smallest offer you would be willing to accept. If the offer the computer generated is above the amount you give \textbf{(Condition 1)}, AND at least 25 buyers match your price \textbf{(Condition 2)}, we will ask you to download your X(Twitter) data, run it through our tool to remove sensitive data, upload it on our marketplace, and pay you the total amount from all matching buyers. If the offer is below that amount, OR fewer than 25 buyers matched your price, we will not ask you to sell (you won’t have another chance to adjust your price). \textbf{Regardless of the match, all participants who complete the study will receive a flat rate of \$1.10 as a thank you for their time.}

\textbf{Example 1:} Computer generated \$0.8. You set a minimum price of \$1, so you don’t pass condition 1. Then, we will pay you \$1.10 for participating in this study. The marketplace would pay you nothing because you failed Condition 1.

\textbf{Example 2:} Computer generated \$0.8. You set a minimum price of \$0.6, so you pass condition 1. Your price is equal or lower than the maximum price of only 20 buyers, which is less than the required minimum of 25 buyers, so you do NOT pass condition 2. Then, we will pay you \$1.10 for participating in this study. The marketplace would pay you nothing because you failed Condition 2.

\textbf{Example 3:} Computer generated \$0.8. You set a minimum price of \$0.6, so you pass Condition 1. Your price is equal or lower than the maximum price of 40 buyers, so you pass Condition 2 as well. Then we will pay you \$1.10 for participating in this study. The marketplace would pay you \$0.6 × 40 = \$24.

\subsection*{Study 2 -- Treatment 3}
\vspace{1ex}
\noindent
\setstretch{1.2}

\textbf{SELL? S2 T3}\\
Are you \textbf{willing to sell} your X(Twitter) data package under these conditions?
\begin{itemize}
    \item Yes
    \item No
\end{itemize}

\textbf{MINPRICE S2 T3}\\
What is \textbf{your minimum price} in two-decimal \$USD format? (e.g. 0.25, 1.00)\\
\hspace*{1em}\textit{Displayed as open textbox if "Yes" was selected in SELL? S2 T3}

\textbf{WILLING S2 T3}\\
In addition to receiving money, is there any other motivation for you in selling your data? (optional)\\
\hspace*{1em}\textit{Displayed as open textbox if "Yes" was selected in SELL? S2 T3}

\textbf{NOTWILLING S2 T3}\\
Can you tell us why you don’t want to sell your data?
\begin{enumerate}
    \item I need to know exactly who will be using my data
    \item I need to understand the specific purpose for which my data is needed
    \item I don't trust X(Twitter)
    \item I don't trust private companies
    \item I trust private companies but am unsure whether they share data with third parties
    \item I am unsure about what is included in my X(Twitter) data
    \item Other
\end{enumerate}

\textbf{S2 T3 OTHER}\\
You have selected "other". Please specify:\\
\hspace*{1em}\textit{Displayed as open textbox if "Other" was selected in NOTWILLING S2 T3}


\subsection*{Study 2 -- Treatment 4 -- Intro}
\vspace{1ex}
\noindent
\setstretch{1.2}

\textbf{S2 T4 INTRO}\\
Social media platforms like X(Twitter) now allow users to export and download their data as a ZIP file. This means all X(Twitter) users can get a full copy of their data. Since X(Twitter) has closed its data access API, small and medium private companies now rely on individuals who are willing to share their data.

We’ve created a marketplace for X(Twitter) data where anyone interested in selling their X(Twitter) data package to small and medium private companies can sign up, set a minimum price, and sell it to \textbf{as many small and medium private companies as possible.} Since launching two weeks ago, we’ve already received offers from buyers who have set their maximum prices. \textbf{All our buyers are small and medium size private companies. We do NOT allow big tech companies like Google and OpenAI to use our marketplace.}

\textbf{Here’s how it works:} Based on the maximum prices the private companies on our marketplace are willing to pay, the computer has randomly generated an amount of money \textbf{between \$0.25 and \$1} to offer you to sell your package of X(Twitter) data to matched companies. Before we tell you what the offer is, we will ask you the smallest offer you would be willing to accept. If the offer the computer generated is above the amount you give \textbf{(Condition 1)}, AND at least 25 buyers match your price \textbf{(Condition 2)}, we will ask you to download your X(Twitter) data, upload it on our marketplace, and pay you the total amount from all matching buyers. If the offer is below that amount, OR fewer than 25 buyers matched your price, we will not ask you to sell (you won’t have another chance to adjust your price). \textbf{Regardless of the match, all participants who complete the study will receive a flat rate of \$1.10 as a thank you for their time.}

\textbf{Example 1:} Computer generated \$0.8. You set a minimum price of \$1, so you don’t pass condition 1. Then, we will pay you \$1.10 for participating in this study. The marketplace would pay you nothing because you failed Condition 1.

\textbf{Example 2:} Computer generated \$0.8. You set a minimum price of \$0.6, so you pass condition 1. Your price is equal or lower than the maximum price of only 20 buyers, which is less than the required minimum of 25 buyers, so you do NOT pass condition 2. Then, we will pay you \$1.10 for participating in this study. The marketplace would pay you nothing because you failed Condition 2.

\textbf{Example 3:} Computer generated \$0.8. You set a minimum price of \$0.6, so you pass Condition 1. Your price is equal or lower than the maximum price of 40 buyers, so you pass Condition 2 as well. Then we will pay you \$1.10 for participating in this study. The marketplace would pay you \$0.6 × 40 = \$24.

\subsection*{Study 2 -- Treatment 4}
\vspace{1ex}
\noindent
\setstretch{1.2}

\textbf{SELL? S2 T4}\\
Are you \textbf{willing to sell} your X(Twitter) data package under these conditions?
\begin{itemize}
    \item Yes
    \item No
\end{itemize}

\textbf{MINPRICE S2 T4}\\
What is \textbf{your minimum price} in two-decimal \$USD format? (e.g. 0.25, 1.00)\\
\hspace*{1em}\textit{Displayed as open textbox if "Yes" was selected in SELL? S2 T4}

\textbf{WILLING S2 T4}\\
In addition to receiving money, is there any other motivation for you in selling your data? (optional)\\
\hspace*{1em}\textit{Displayed as open textbox if "Yes" was selected in SELL? S2 T4}

\textbf{NOTWILLING S2 T4}\\
Can you tell us why you don’t want to sell your data?
\begin{enumerate}
    \item I need to know exactly who will be using my data
    \item I need to understand the specific purpose for which my data is needed
    \item I don't trust X(Twitter)
    \item I don't trust private companies
    \item I trust private companies but am unsure whether they share data with third parties
    \item I am unsure about what is included in my X(Twitter) data
    \item Other
\end{enumerate}

\textbf{S2 T4 OTHER}\\
You have selected "other". Please specify:\\
\hspace*{1em}\textit{Displayed as open textbox if "Other" was selected in NOTWILLING S2 T4}


\subsection*{Study 2 -- Control -- Intro}
\vspace{1ex}
\noindent
\setstretch{1.2}

\textbf{S2 CONTROL INTRO}\\
Social media platforms like X(Twitter) now allow users to export and download their data as a ZIP file. This means all X(Twitter) users can get a full copy of their data. Since X(Twitter) has closed its data access API, researchers now rely on individuals who are willing to donate their data.

We’ve created a tool where anyone interested in donating their complete X(Twitter) data package to us for research can sign up and send us their X(Twitter) ZIP file through a secure channel. \textbf{Our research project received ethics approval from our university’s ethics committee.}

\textbf{All participants who complete the study will receive a flat rate of \$1.10 as a token of appreciation, regardless of whether they choose to donate their data.}

\subsection*{Study 2 -- Control}
\vspace{1ex}
\noindent
\setstretch{1.2}

\textbf{DONATE S2 CONTROL}\\
Are you \textbf{willing to donate} your X(Twitter) data package to us \textbf{for free} under these conditions?\\
(All participants who complete the study will receive a flat rate of \$1.10 as a token of appreciation, regardless of whether they choose to donate their data.)
\begin{itemize}
    \item Yes
    \item No
\end{itemize}

\textbf{WILLING S2 CONTROL}\\
Can you tell us about your motivation for donating your data? (optional)\\
\hspace*{1em}\textit{Displayed as open textbox if "Yes" was selected in DONATE S2 CONTROL}

\textbf{NOTWILLING S2 CONTROL}\\
Can you tell us why you don’t want to donate your data?
\begin{enumerate}
    \item I need to know exactly who will be using my data
    \item I need to understand the specific purpose for which my data is needed
    \item I don't trust X(Twitter)
    \item I don't trust researchers
    \item I trust researchers but am unsure whether they share data with third parties
    \item I am unsure about what is included in my X(Twitter) data
    \item Other
\end{enumerate}

\textbf{S2 CONTROL OTHER}\\
You have selected "other". Please specify:\\
\hspace*{1em}\textit{Displayed as open textbox if "Other" was selected in NOTWILLING S2 CONTROL}


\subsection*{Study 2 -- Post-Survey}
\vspace{1ex}
\noindent
\setstretch{1.2}

\textbf{POST-SURVEY-INTRO}\\
In the following questions, we are curious to find out your knowledge on online data privacy and security.

\textbf{S2.1}\\
Please indicate below to which degree you agree/disagree with each of the following statements.\\
\hspace*{1em}\textit{Scale: 1 = Strongly disagree, 2 = Somewhat disagree, 3 = Somewhat agree, 4 = Strongly agree}
\begin{itemize}
    \item Consumers have lost all control over how personal information is collected and used by companies.
    \item Most businesses handle the personal information they collect about consumers in a proper and confidential way.
    \item Existing laws and organizational practices provide a reasonable level of protection for consumer privacy today.
\end{itemize}

\textbf{S2.2.1-1}\\
Have you ever blocked anyone on your X(Twitter) account?
    \begin{itemize}
        \item Yes
        \item No
    \end{itemize}

\textbf{S2.2.1-2}\\
Have you ever muted anyone on your X(Twitter) account?
    \begin{itemize}
        \item Yes
        \item No
    \end{itemize}

\textbf{S2.2.1-3}\\
Have you ever reported any X(Twitter) account?
    \begin{itemize}
        \item Yes
        \item No
    \end{itemize}

\textbf{S2.2.1-4}\\
Have you experienced previous privacy violations?
    \begin{itemize}
        \item No
        \item Yes, once
        \item Yes, more than once
        \item I don't know
    \end{itemize}

\textbf{S2.2.2}\\
Please indicate below to which degree you agree/disagree with each of the following statements.\\
\hspace*{1em}\textit{Scale: 1 = Strongly disagree, 2 = Disagree, 3 = Somewhat disagree, 4 = Neither agree nor disagree, 5 = Somewhat agree, 6 = Agree, 7 = Strongly agree}
\begin{itemize}
    \item I have the knowledge necessary to use privacy features to regulate information on X(Twitter).
    \item Given the knowledge it takes to use the privacy features, it would be easy for me to control information flow on X(Twitter).
\end{itemize}

\textbf{S2.2.3 \& S2.2.4}\\
How frequently do you use privacy features on X(Twitter)?\\
\hspace*{1em}\textit{Scale: 1 = Never, 2 = Rarely, 3 = Occasionally, 4 = Sometimes, 5 = Frequently, 6 = Usually, 7 = Always}
\begin{itemize}
    \item How frequently do you post to the selected group of friends using privacy features (e.g., Twitter lists)?
    \item How frequently do you post a status update excluding some friends?
    \item How frequently have you deleted posts that others made on your timeline?
    \item How frequently have you untagged yourself in a photo or post that was posted by others?
\end{itemize}

\subsection*{Study 2 -- Debriefing}
\vspace{1ex}
\noindent
\setstretch{1.2}
Thank you for your participation in this research study. Now that you completed or have ended your participation, we will provide you with some additional information about the purposes of this study.

\textbf{What you should know about this study:} 
The main goal of this study was to see how likely people are to sell all their X(Twitter) data and to understand the minimum price they would set for it. We haven’t actually created a marketplace yet, so we won’t be asking you to download or send your X(Twitter) data. We mentioned having such a platform to help create a realistic experience for participants.

\textbf{If you have questions:} 
The main researchers conducting this study are \textit{(blinded for peer review)}. If you have questions, you may contact one of the main researchers. If you have any questions or concerns regarding your rights as a research participant in this study, you may contact the Ethics Committee: 
\textit{(contact info Ethics Committee)}

\setcounter{figure}{0} \renewcommand{\thefigure}{S\arabic{figure}}
\setcounter{table}{0} \renewcommand{\thetable}{S\arabic{table}}

\clearpage

\end{document}